\documentclass{aa}
\usepackage[varg]{txfonts}
\usepackage{multirow}
\usepackage{natbib,twoopt}
\usepackage{amsmath}
\usepackage{mathtools}
\usepackage{xr}
\usepackage{soul}
\usepackage[breaklinks=true]{hyperref} 
\bibpunct{(}{)}{;}{a}{}{,}             
\makeatletter
  \newcommandtwoopt{\citeads}[3][][]{\href{http://adsabs.harvard.edu/abs/#3}%
    {\def\hyper@linkstart##1##2{}%
     \let\hyper@linkend\@empty\citealp[#1][#2]{#3}}}
  \newcommandtwoopt{\citepads}[3][][]{\href{http://adsabs.harvard.edu/abs/#3}%
    {\def\hyper@linkstart##1##2{}%
     \let\hyper@linkend\@empty\citep[#1][#2]{#3}}}
  \newcommandtwoopt{\citetads}[3][][]{\href{http://adsabs.harvard.edu/abs/#3}%
    {\def\hyper@linkstart##1##2{}%
     \let\hyper@linkend\@empty\citet[#1][#2]{#3}}}
  \newcommandtwoopt{\citeyearads}[3][][]%
    {\href{http://adsabs.harvard.edu/abs/#3}
    {\def\hyper@linkstart##1##2{}%
     \let\hyper@linkend\@empty\citeyear[#1][#2]{#3}}}
\usepackage{etoolbox}
\makeatletter
\patchcmd\@combinedblfloats{\box\@outputbox}{\unvbox\@outputbox}{}{%
    \errmessage{\noexpand\@combinedblfloats could not be patched}%
}%
\makeatother

\newcommand{\BIZt}{\citetads{2014A&A...569A..27B}}

\newcommand{\SPEt}{\citetads{2016A&A...592L..11S}}
\newcommand{\SPEp}{\citepads{2016A&A...592L..11S}}

\begin{document}
\title{Mapping deuterated methanol toward L1544: I. Deuterium fraction and comparison with modeling \thanks{Based on observations carried out with the IRAM 30 m Telescope. IRAM is supported by INSU/CNRS (France), MPG (Germany) and IGN (Spain)} }

\author{A.~Chac\'on-Tanarro \inst{1}
\and P.~Caselli \inst{1}
\and L.~Bizzocchi \inst{1}
\and J.~E.~Pineda \inst{1}
\and O.~Sipil\"a \inst{1}
\and A.~Vasyunin \inst{1,2}
\and S.~Spezzano \inst{1}
\and A.~Punanova \inst{2,1}
\and B.M.~Giuliano \inst{1}
\and V.~Lattanzi \inst{1}
}
\institute{Max-Planck-Instit\"{u}t f\"{u}r extraterrestrische Physik, Giessenbachstrasse 1, 85748 Garching, Germany
\and Ural Federal University, 620002, 19 Mira street, Yekaterinburg, Russia}
\date{Received - / Accepted -}

\abstract{The study of deuteration in pre-stellar cores is important to understand the physical and chemical initial conditions in the process of star formation. In particular, observations toward pre-stellar cores of methanol and deuterated methanol, solely formed on the surface of dust grains, may provide useful insights on surface processes at low temperatures.}{ Here we analyze maps of CO, methanol, formaldehyde and their deuterated isotopologues toward a well-known pre-stellar core. This study allows us to test current gas-dust chemical models.}{Single-dish observations of CH$_3$OH, CH$_2$DOH, H$_2$CO,  H$_2\,^{13}$CO, HDCO, D$_2$CO and C$^{17}$O toward the prototypical pre-stellar core L1544 were performed at the IRAM 30 m telescope. We analyze their column densities, distributions, and compare these observations with gas-grain chemical models. } {The maximum deuterium fraction derived for methanol is [CH$_2$DOH]/[CH$_3$OH] $\sim$ 0.08$\pm$0.02, while the measured deuterium fractions of formaldehyde at the dust peak are [HDCO]/[H$_2$CO] $\sim$ 0.03$\pm$0.02, [D$_2$CO]/[H$_2$CO] $\sim$ 0.04$\pm$0.03 and [D$_2$CO]/[HDCO] $\sim$ 1.2$\pm$0.3. Observations differ significantly from the predictions of models, finding discrepancies between a factor of 10 and a factor of 100 in most cases. It is clear though that to efficiently produce methanol on the surface of dust grains, quantum tunneling diffusion of H atoms must be switched on. It also appears that the currently adopted reactive desorption efficiency of methanol is overestimated and/or that abstraction reactions play an important role. More laboratory work is needed to shed light on the chemistry of methanol, an important precursor of complex organic molecules in space.}{}


\maketitle

\section{Introduction}

Pre-stellar cores form in molecular clouds, due to the influence of gravity, magnetic fields and turbulence. They are dense ($n_{\rm{H}_2}>10^5$\,cm$^{-3}$) and cold ($T<$10\,K) toward their center \citepads{2010MNRAS.402.1625K}. They are the starting point in the process of star formation \citepads{2007ARA&A..45..339B, 2012A&ARv..20...56C}, as they are self-gravitating dense cores which present signs of contraction motions and chemical evolution \citepads{2005ApJ...619..379C}. Hence, they represent ideal laboratories to study the early evolutionary stages of low-mass star formation.

Methanol (CH$_3$OH) is one of the most widespread organic molecules in the ISM and a major precursor of chemical complexity in space. Methanol is believed to form on dust grain surfaces by sequential addition of hydrogen atoms to adsorbed CO molecules \citepads{1982A&A...114..245T, 2002ApJ...571L.173W}. This process should take place in cold dense cores at densities above 10$^4$\,cm$^{-3}$, where large amounts of gas-phase CO molecules start to freeze-out onto dust grain surfaces \citepads{1998ApJ...507L.171W, 1999ApJ...523L.165C, 2002ApJ...569..815T}. 

In L1544, \citetads{2014ApJ...795L...2V} deduced that the CH$_3$OH emission should trace an external layer of the core. In fact, \citetads{2014A&A...569A..27B} mapped the CH$_3$OH emission across L1544, finding an asymmetric ring-like shape surrounding the dust peak. The methanol peak is about 4000 au away from the core center and its distribution can be reproduced by the recent gas-grain chemical model (applied to L1544) by \citetads{2017ApJ...842...33V}. This can be understood by the fact that the CH$_3$OH molecule can more easily desorb from the grain surface when a significant fraction of surface ice is in CO, and CO-rich dust surfaces are present around 4000-5000 au away from the core center because of the fast CO freeze-out at that location and the different CO and H$_2$O photodesorption rates \citepads{2017ApJ...842...33V}.

Deuterated methanol should follow the methanol distribution, as they are both formed on the surface of dust grains. However, as already found by previous authors \citepads{2002ApJ...565..344C}, the deuterium fraction increases toward the densest and coldest parts of the core, where CO is mainly in solid form. Toward the center, due to the dissociative recombination of the abundant deuterated forms of H$_3^+$, the D/H atomic ratio increases from the cosmic value \citepads[$\sim$1.5$\times$10$^{-5}$; ]{2006ApJ...647.1106L} to values larger than 0.1 \citepads[e.g.]{2003ApJ...591L..41R}. Thus, toward the core center, D atoms compete with H atoms in the saturation of CO molecules, efficiently producing deuterated methanol
 \citepads{2012ApJ...760...40A,2012ApJ...748L...3T,2014prpl.conf..859C}. Observations toward L1544 showed that the centroid velocity ($V_{LSR}$) of CH$_2$DOH is about 0.2 km/s lower than that of CH$_3$OH \citepads{2014A&A...569A..27B}; this shift is the same as the one found when comparing the $V_{LSR}$ of C$^{17}$O (1-0), which traces the outer parts of the core, with the $V_{LSR}$ of N$_2$D$^+$ (2-1), mainly tracing the dense central regions where CO is frozen out \citepads{2002ApJ...565..331C}, thus suggesting that CH$_2$DOH is tracing denser regions compared to CH$_3$OH, in agreement with the theoretical expectations.  However, this hypothesis is based on the results of single pointing observations, so no information on the spatial distribution of CH$_2$DOH in the core can be inferred. 

Formaldehyde (H$_2$CO) can be formed both via grain--surface chemistry (in an intermediate step of the formation of methanol) and gas-phase chemistry \citepads[e.g. through reactions of hydrocarbons with oxygen atoms;]{2017iace.book.....Y}. Which route actually dominates is still a matter of debate \citepads{2006A&A...453..949P}. In fact, \citetads{2009A&A...508..737P} suggested that pure gas-phase chemistry can account for the abundances of formaldehyde and its deuterated species observed toward the Orion Bar PDR. On the other hand, \citetads{2011A&A...527A..39B}  had to invoke grain--surface chemistry to explain the high deuteration found in the $\rho$ Oph~A cloud. Interestingly, these authors found D$_2$CO to be more abundant than HDCO. 

The deuterium fraction measured in N$_2$H$^+$, NH$_3$, H$_2$CO and CH$_3$OH \citepads{2009A&A...493...89E, 2017MNRAS.467.3011B} reaches the largest values in dynamically evolved starless cores (pre-stellar cores) and toward the youngest protostellar objects, while it decreases in more evolved phases in the low-mass star formation process. The study of different evolutionary stages is thus important to be able to understand how the deuterium budget builds up in molecules formed preferentially in the gas phase and on dust grain surfaces, and to follow these processes during the dynamical evolution of star forming clouds.
We focus our investigation on L1544, a nearby well-known pre-stellar core in the Taurus Molecular Cloud.
Here, we study the spatial features of the methanol and formaldehyde deuteration, together with the CO distribution, as CO is thought to be the parent molecule of methanol. The objective is to gain insights on the surface chemistry and the early history of deuteration during the formation of low-mass stars, which is still unknown.

The structure of the Chapter is the following: in Section \ref{Observations2} we describe the observations, followed by the results in Section \ref{results2}. The analysis, comparison with models, and discussion are presented in Sections \ref{analysis2}, \ref{models2} and \ref{discussion2}, respectively. Our findings are finally summarized in Section \ref{conclusions2}.

\section{Observations} \label{Observations2}

The data presented here were observed using the IRAM 30\,m telescope, located at Pico Veleta (Spain), during 4 observing sessions in 2014 and 2015. We obtained On-The-Fly (OTF) maps for 3 and 2 mm emission lines of the molecules: CH$_2$DOH, CH$_3$OH, H$_2$CO, H$_2\,^{13}$CO, HDCO, D$_2$CO, and C$^{17}$O (see Table \ref{table1}). A region of 2.5\arcmin\ $\times$ 2.5\arcmin\ was mapped in all the lines, except C$^{17}$O, for which a  4\arcmin\ $\times$ 4\arcmin\ OTF map was performed. We applied beam switching for these maps. The pointing accuracy is $<$4\arcsec.
Various E090/E150 EMIR (Eight MIxer Receiver, heterodyne receiver) setups were adopted, and VESPA (VErsatile SPectrometer Assembly) was used as backend for CH$_3$OH, CH$_2$DOH, and C$^{17}$O, with a spectral resolution of 20\,kHz. The data for formaldehyde and its deuterated species were collected using the FTS backend with a spectral resolution of 50\,kHz. All the lines of CH$_3$OH and H$_2$CO were observed simultaneously; CH$_2$DOH (2$_{0,2}$-1$_{0,1}$), CH$_2$DOH (3$_{0,3}$-2$_{0,2}$), C$^{17}$O (1-0), HDCO (2$_{1,1}$-1$_{1,0}$) and D$_2$CO (2$_{1,2}$-1$_{1,1}$) were observed in other different set-ups.
All the maps were finally convolved to a common angular resolution of 30\arcsec, in order to facilitate the comparison between different molecules.  The initial data reduction was done with GILDAS\footnote{https://www.iram.fr/IRAMFR/GILDAS/}.

\begin{figure*}
\resizebox{\hsize}{!}{\includegraphics{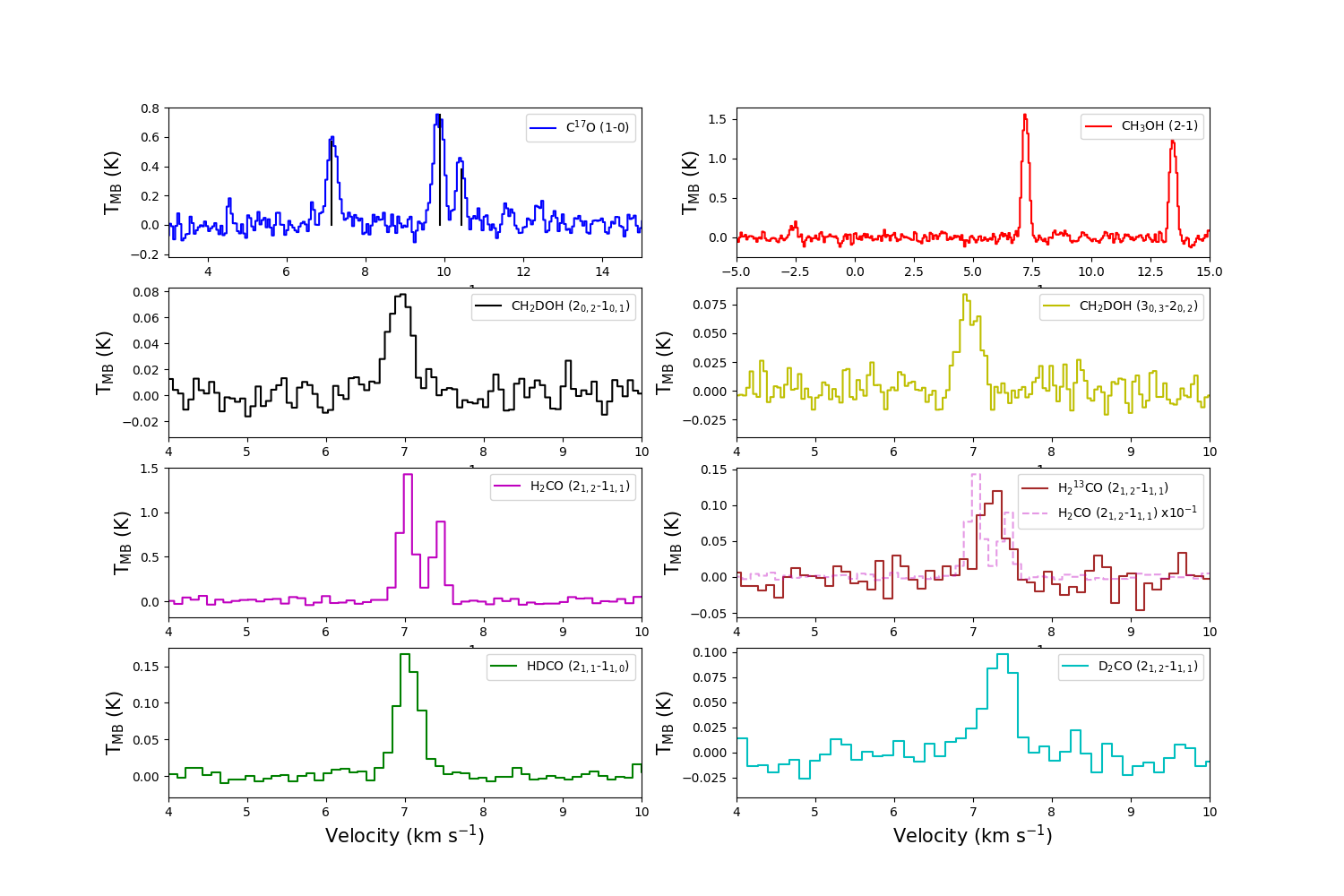}} 
\caption[Spectra of the lines C$^{17}$O (1-0), CH$_3$OH (2-1), CH$_2$DOH (2$_{0,2}$-1$_{0,1}$, e$_0$), CH$_2$DOH (3$_{0,3}$-2$_{0,2}$, e$_0$), H$_{2}$CO (2$_{1,2}$-1$_{1,1}$), H$_2\,^{13}$CO (2$_{1,2}$-1$_{1,1}$), HDCO (2$_{1,1}$-1$_{1,0}$), and D$_2$CO (2$_{1,2}$-1$_{1,1}$) at the center of L1544]{Averaged spectra of all lines in the central $\sim$30$\times$30 arcsec$^2$. The specifications of the lines and the spectra are shown in Table \ref{table1}. The H$_2$CO (2$_{1,2}$-1$_{1,1}$) spectrum is superposed on the H$_2\,^{13}$CO (2$_{1,2}$-1$_{1,1}$) spectrum to show that the H$_2$CO (2$_{1,2}$-1$_{1,1}$) line is self-absorbed. Superposed on the C$^{17}$O spectrum there are three lines that indicate the relative intensities of the hyperfine components.} 
\label{averaged_spectra}
\end{figure*}

The observations were done under average good weather conditions ($\tau_{225\,\mathrm{GHz}}\sim0.1-0.5$, i.e. $pwv\sim2-8$mm), and with $T_{\mathrm{sys}}$ ranging from $\sim$100 to $\sim$150 K for the 3 mm band, and from $\sim$100 K to $\sim$160 K for the 2 mm band. All spectra were converted from $T_A^*$ to $T_{mb}$ using the forward and main beam efficiency ratios shown in Table \ref{table1}, due to the extended but not uniform nature of the maps here presented. We note that this may overestimate our results by $\sim$20\% for the more extended emission of C$^{17}$O (1-0), but this is within the uncertainties and it does not change the conclusions derived from the morphology of the emission distribution of C$^{17}$O (1-0) across the map (see Section \ref{results2}). 
The spectra obtained averaging the central 30$\times$30 arcsec$^{2}$ of each map are shown in Fig.~\ref{averaged_spectra}.

\begin{table*}
\caption{Lines observed, their rest frequency, noise level, velocity resolution, map pixel size, forward and beam efficiency ratio, and the references for the spectroscopic information for each line.}
\label{table1}
\begin{tabular}{ccccccc}
\hline
\hline
Line & Rest frequency & rms & Velocity resolution & Pixel size & $F_{eff}/B_{eff}$ & References \\
 & (MHz) & (mK) & (km\,s$^{-1}$) & (arcsec) & &  \\
\hline
CH$_3$OH (2$_{0,2}$-1$_{0,1}$, A$^+$) & \,\,\,96\,741.375 $\pm$ 0.003 & 62 & 0.06 & 5 & 1.18 & 1  \\ 
CH$_3$OH (2$_{1,2}$-1$_{1,1}$, E$_2$) & \,\,\,96\,739.362 $\pm$ 0.003 & 62 & 0.06 & 5 & 1.18 & 1 \\
CH$_3$OH (2$_{0,2}$-1$_{0,1}$, E$_1$) & \,\,\,96\,744.550 $\pm$ 0.003 & 62 & 0.06 & 5 & 1.18 & 1  \\

CH$_2$DOH (2$_{0,2}$-1$_{0,1}$, e$_0$) & \,\,\,89\,407.817 $\pm$ 0.002 & 15 & 0.07 & 5 & 1.18 &  2 \\
CH$_2$DOH (3$_{0,3}$-2$_{0,2}$, e$_0$) & 134\,065.381 $\pm$ 0.002 & 15 & 0.04 & 5 & 1.25 & 2 \\

H$_{2}$CO (2$_{1,2}$-1$_{1,1}$) & 140\,839.502 $\pm$ 0.010 & 40 & 0.10 & 5 & 1.26 & 3, 4 \\

H$_2\,^{13}$CO (2$_{1,2}$-1$_{1,1}$) & 137\,449.950 $\pm$ 0.004 & 30 & 0.11 & 5 & 1.26 & 5 \\

HDCO (2$_{1,1}$-1$_{1,0}$) & 134\,284.830 $\pm$ 0.100 & 10 & 0.11 & 5 & 1.20 & 6 \\

D$_2$CO (2$_{1,2}$-1$_{1,1}$) & 110\,837.830 $\pm$ 0.100 & 17 & 0.13 & 5 & 1.25 & 6 \\

C$^{17}$O (1-0) & 112\,360.007 $\pm$ 0.015 & 75 & 0.05 & 4 & 1.20 & 7 \\
\hline
 \end{tabular}
 
 \raggedright
References: (1) \citetads{1997JPCRD..26...17X}; (2) \citetads{2012mss..confERF12P}; (3) \citetads{MULLER201728}; (4)  \citetads{CORNET1980438}; (5) \citetads{B002819N}; (6) \citetads{BOCQUET1999345}; (7) \citetads{2003ApJ...582..262K}. The references and values from (3) to (7) were found using the CDMS \citepads[Cologne Database for Molecular Spectroscopy;]{2001A&A...370L..49M, 2005JMoSt.742..215M}.

\end{table*}

\section{Results}\label{results2}

The integrated intensities were derived calculating the zeroth moment of the maps using a custom code written in Python, making use of Astropy \citepads{2013A&A...558A..33A}, SciPy and NumPy \citepads{scipi}. In this Section only part of the maps is presented. For the rest of the integrated intensity maps, see Appendix \ref{appendix_intensity}. The kinematic analysis will be presented in an upcoming paper (Chac\'on-Tanarro et al., in prep., Paper II).
\begin{itemize}
\item{CH$_3$OH and CH$_2$DOH}

In Fig.~\ref{moment0_ch3oh-ch2doh}, the integrated intensity map of the transitions A$^+$ (2$_{0,2}$-1$_{0,1}$) of methanol and (2$_{0,2}$-1$_{0,1}$) and (3$_{0,3}$-2$_{0,2}$) of deuterated methanol are shown. Methanol, as previously reported, shows an asymmetric ring-like structure around the center of the core peaking toward the north-east at a distance of 4000 au. In fact, when the emission seen in methanol is averaged in concentric ellipses, as previously done for the millimeter continuum maps in \citetads{2017A&A...606A.142C}, it shows CH$_3$OH depleted towards the center and enhanced at a distance of $\sim$3000\,au, following a ring-like shape. The emission peak is shifted towards a more central region during this process (by $\sim$1000\,au). This shift is with respect to the position of the methanol peak (as seen in Fig. \ref{moment0_ch3oh-ch2doh}), as the annular average includes regions with fainter CH$_3$OH emission, which reduces the size of the ring compared to the distance of the CH$_3$OH peak from the dust peak. The asymmetric distribution and the faint emission towards the South was already noticed by \citetads{2016A&A...592L..11S}, who showed that the H$_2$ column density has a sharp drop toward the Southern part of the core, where in fact photoprocesses dominate. The peak of CH$_2$DOH map extends toward the dust peak, suggesting that deuterated methanol may trace a higher density zone compared to the main isotopologue.  We note that the clumpiness seen in the CH$_2$DOH (2$_{0,2}$-1$_{0,1}$) map is due to noise (the noise in the integrated intensity is 0.005 K\,km\,s$^{-1}$), and that the distribution of CH$_2$DOH is better seen with the map of the (3$_{0,3}$-2$_{0,2}$) line. 

\begin{figure*}
\includegraphics[width=1.0\textwidth]{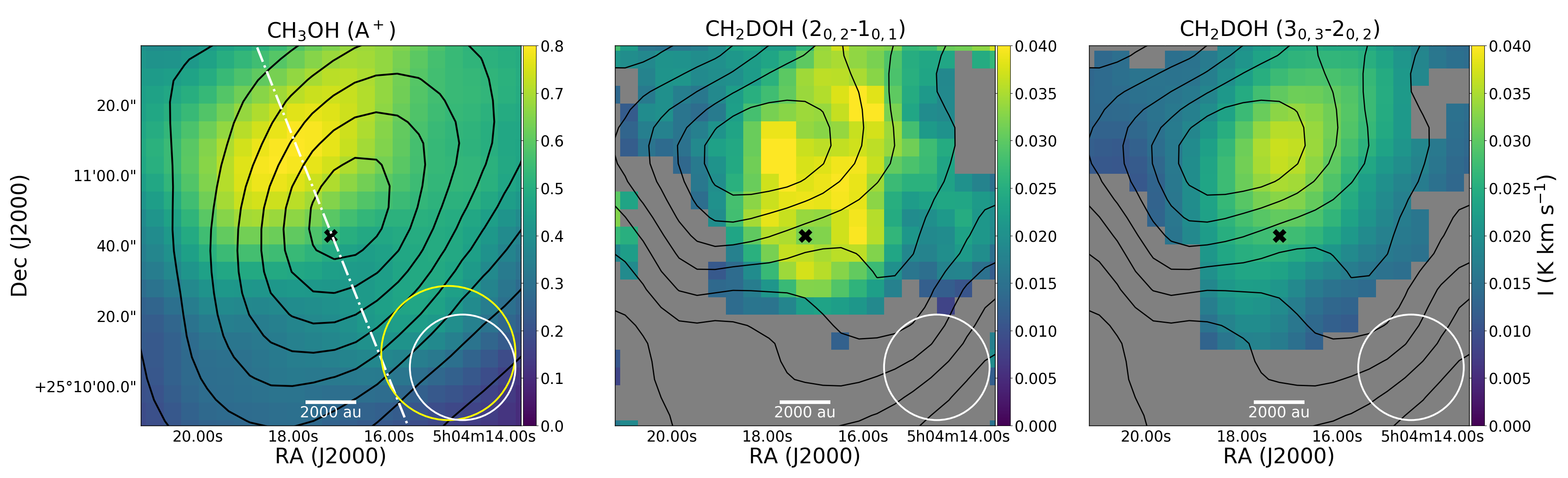}
\caption[Integrated intensity maps of the lines CH$_3$OH (2$_{0,2}$-1$_{0,1}$, A$^+$), CH$_2$DOH  (2$_{0,2}$-1$_{0,1}$), and CH$_2$DOH (3$_{0,3}$-2$_{0,2}$)]{\textit{Left panel:} integrated intensity map of the strongest methanol transition observed (2$_{0,2}$-1$_{0,1}$, A$^+$). The black contours represent increasing 10\% steps of the N$_{\rm{H}_2}$ column density map, derived by \citetads{2016A&A...592L..11S} using Herschel/SPIRE data. The noise in the integrated intensity is 0.02 K km s$^{-1}$. The white dashed-dotted line indicates the cut used for comparison with the models (see Section~\ref{models2}). \textit{Middle panel:} integrated intensity map of the deuterated methanol (2$_{0,2}$-1$_{0,1}$) transition. The black contours represent increasing 10\% steps of the CH$_3$OH column density map, derived as explained in Section~\ref{analysis2}. The error in the integrated intensity is 0.005 K km s$^{-1}$.  \textit{Right panel: }integrated intensity map of the deuterated methanol (3$_{0,3}$-2$_{0,2}$) transition. The black contours represent increasing 10\% steps of the CH$_3$OH column density map, derived as explained in Section~\ref{analysis2}. The error in the integrated intensity is 0.004 K km s$^{-1}$. In all panels the HPBWs are shown in the bottom right corner of the figure, in yellow for \textit{Herschel}/SPIRE and in white for the 30 m telescope,  the black cross marks the dust continuum peak and only pixels with detection level above 3$\sigma$ are included. } 
\label{moment0_ch3oh-ch2doh}
\end{figure*}

\item{H$_2$CO, H$_2\,^{13}$CO, HDCO and D$_2$CO}

The integrated intensity maps of H$_2\,^{13}$CO (2$_{1,2}$-1$_{1,1}$), HDCO (2$_{1,1}$-1$_{1,0}$) and D$_2$CO (2$_{1,2}$-1$_{1,1}$) are shown in Fig.~\ref{moment0_hdco-d2co}, and the integrated intensity map of H$_2$CO (2$_{1,2}$-1$_{1,1}$) is shown in Fig.~\ref{moment0_h2co}. The H$_2$CO line is asymmetric and double-peaked, with the blue peak stronger than the red peak (see Fig.~\ref{averaged_spectra}), suggestive of contraction motions, as already seen by \citetads{1998ApJ...504..900T} and \citetads{2003ApJ...585L..55B}. The H$_2\,^{13}$CO (2$_{1,2}$-1$_{1,1}$) line matches with the absorption dip from the H$_2$CO (2$_{1,2}$-1$_{1,1}$) line, indicating that this feature is indeed due to self-absorption (see Fig.~\ref{averaged_spectra}). Thus, the distribution of formaldehyde is better traced by H$_2\,^{13}$CO, which also presents an ring-like structure around the core center, with a peak close to the methanol peak and, unlike CH$_3$OH, a secondary maximum toward the South. However, HDCO and D$_2$CO show centrally concentrated emission, differing from CH$_3$OH, CH$_2$DOH and H$_2$CO. 

\begin{figure*}[h]
\resizebox{\hsize}{!}{\includegraphics{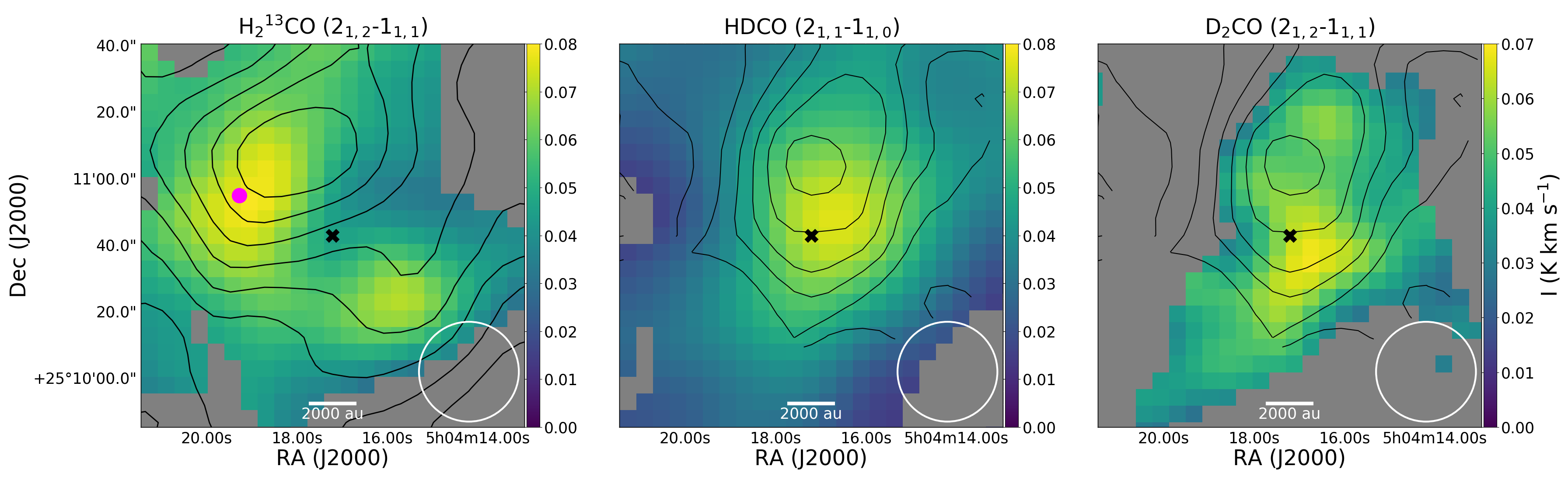}} 
\caption[Integrated intensity  maps of the lines H$_2\,^{13}$CO (2$_{2,1}$-1$_{1,1}$), HDCO (2$_{1,1}$-1$_{1,0}$), and D$_2$CO (2$_{1,2}$-1$_{1,1}$)]{\textit{Left panel:} integrated intensity map of H$_2\,^{13}$CO (2$_{2,1}$-1$_{1,1}$). The error in the integrated intensity is 0.01 K km s$^{-1}$. The black contours represent increasing 10\% steps of the CH$_3$OH column density map, derived as explained in Section~\ref{analysis2}. The pink circle marks the C$^{17}$O peak (see also Fig.~\ref{moment0_c17o}). \textit{Middle panel: }integrated intensity maps of HDCO (2$_{1,1}$-1$_{1,0}$). The error in the integrated intensity is 0.005 K km s$^{-1}$. The black contours represent 10\% steps with respect to the deuterated methanol column density peak. \textit{Right panel:} integrated intensity map of D$_2$CO (2$_{1,2}$-1$_{1,1}$). The error in the integrated intensity is 0.01 K km s$^{-1}$. The black contours represent 10\% steps with respect to the deuterated methanol column density peak. In all panels, the HPBW is shown in the bottom right corners, and the black cross marks the dust continuum peak. Only pixels with detection level above 3$\sigma$ are included.}
\label{moment0_hdco-d2co}
\end{figure*}

\item{C$^{17}$O}

The integrated intensity of C$^{17}$O (1-0) is shown in Fig.~\ref{moment0_c17o}. For the integrated intensity and the column density of C$^{17}$O, only the isolated component of the hyperfine structure (F=$\frac{5}{2}$-$\frac{5}{2}$) is used. The distribution of C$^{17}$O looks similar to the previous map presented by \citetads{1999ApJ...523L.165C}, although here the emission peak is more pronounced and the distribution is clumpier. These differences are caused by the different mapping methods, as \citetads{1999ApJ...523L.165C} only observed selected points while we carried out on-the-fly mapping of a more extended region. Interestingly, the emission peak coincides with the peak seen in H$_2\,^{13}$CO, as shown in Fig.~\ref{moment0_hdco-d2co}.  We point out that the C$^{17}$O (1-0) transition is optically thin across the core, as already found by \citetads{1999ApJ...523L.165C}; this can also be seen from Fig.~\ref{averaged_spectra}, where the relative intensities of the hyperfine components have been indicated below the observed spectrum.

\begin{figure*}
\resizebox{\hsize}{!}{\includegraphics{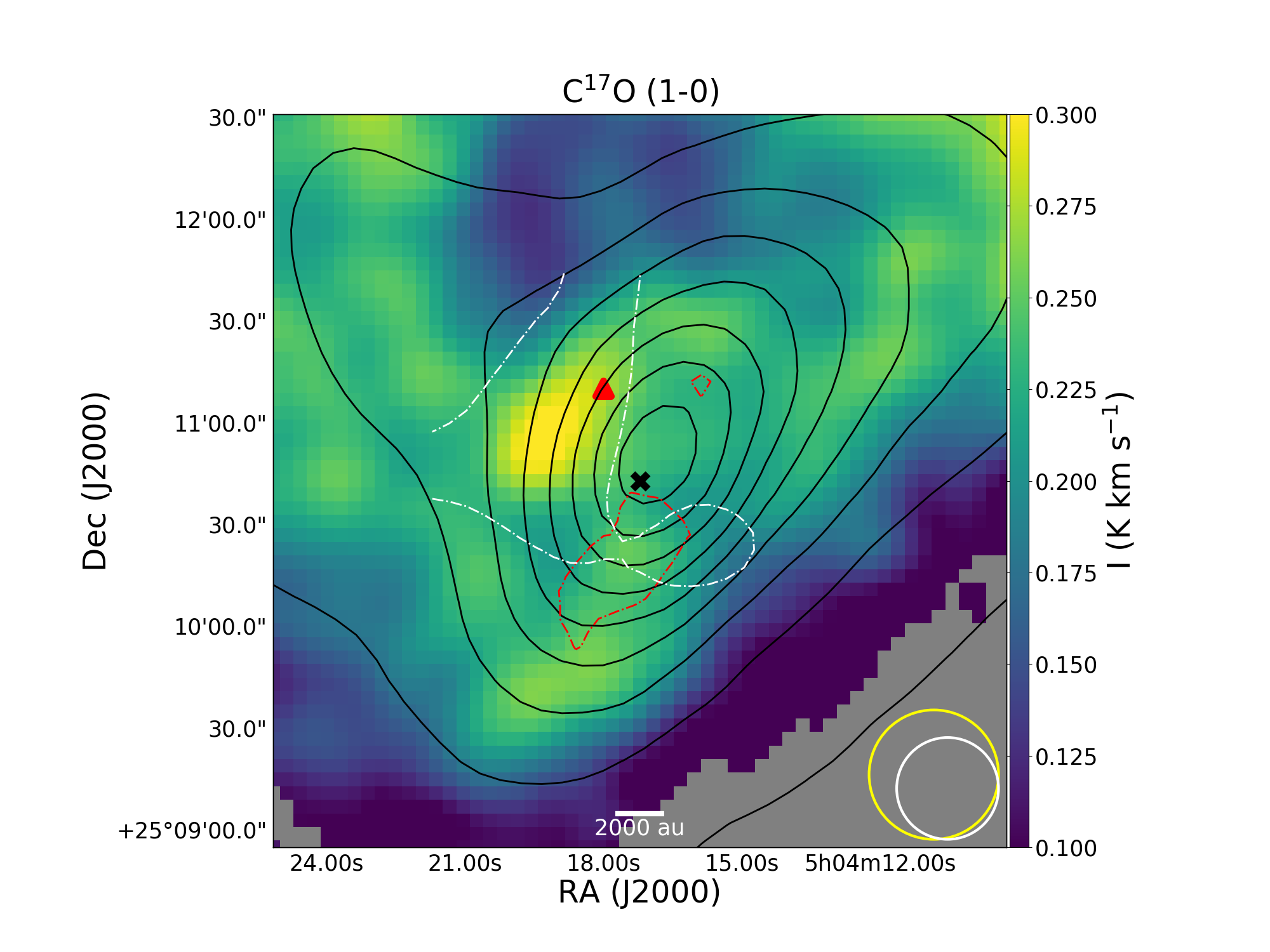}} 
\caption[Integrated intensity map of the line C$^{17}$O (1-0)]{Integrated intensity map of the C$^{17}$O (1-0) line. The error in the integrated intensity is 0.02 K km s$^{-1}$. The black contours represent increasing 10\% steps of the peak of the Herschel N(H$_2$) map, presented by \SPEt\, the white dash-dotted contour shows the 70\% level of the H$_2\,^{13}$CO (2$_{2,1}$-1$_{1,1}$) column density map, derived as explained in Section \ref{cdens}, and the red dash-dotted contour shows the 14$\sigma$ level ($\sigma=$\,7 mK\,km\,s$^{-1}$) of the integrated intensity of the $c$-C$_3$H$_2$ (3$_{2,2}$-3$_{1,3}$) line \SPEp\. The HPBWs are shown in the bottom right corner, in yellow for \textit{Herschel}/SPIRE and in white for the 30 m telescope. The black cross marks the dust continuum peak and the red triangle marks the methanol peak. } 
\label{moment0_c17o}
\end{figure*}

\end{itemize}

\section{Analysis}\label{analysis2}

\subsection{Column densities \label{cdens}}
All the column densities (see Section \ref{h2co} for H$_2$CO) were derived assuming optically thin emission and assuming constant excitation temperature for all levels, following \citepads{2002ApJ...565..344C}:

\begin{equation}\label{eqcolumn}
N = \frac{8\pi\nu^{3}}{c^3}\frac{Q(T_{ex})}{g_uA_{ul}}\left[J_{\nu}(T_{ex})-J_{\nu}(T_{bg})\right]^{-1} \frac{\mathrm{e}^{\frac{E_u}{kT_{ex}}}}{\mathrm{e}^{\frac{h\nu}{kT_{ex}}}-1} \int T_{mb}\mathrm{d}v ,
\end{equation}
$Q(T_{ex})$ being the partition function of the molecule at an excitation temperature $T_{ex}$, $g_u$  the rotational degeneracy of the upper level, $A_{ul}$ the Einstein coefficient for spontaneous emission, $E_u$ the energy of the upper level, $J_{\nu}(T)$ the Rayleigh-Jeans Equivalent Temperature at the frequency $\nu$ and temperature $T$, and $T_{bg}$ the temperature of the background, here the cosmic background (2.7 K). 

\citetads{2014A&A...569A..27B} obtained an excitation temperature for methanol of $T_{ex}$\,=\,6\,$\pm$\,3\,K, and derived the column density of deuterated methanol averaging the results for three different excitation temperatures within the same temperature range: 5, 6.5 and 8 K. The excitation temperature of CH$_2$DOH  cannot be derived because only two transitions were observed. Therefore, they calculated three column densities for CH$_2$DOH using the three excitation temperatures to have an estimate of the uncertainty. Similarly, we derive the column density for both deuterated methanol lines assuming those three different excitation temperatures and averaging all the resulting column densities for the two lines. For methanol, the column density was derived in the same way using the two strongest lines observed (A$^+$ and E$_2$). \BIZt \, derived an optical depth of $\tau$<0.4, and already demonstrated that considering optical depth effects does not vary significantly the resulting column density; thus, the assumption of optically thin emission is justified.

The values found for the column densities of CH$_2$DOH and CH$_3$OH are slightly higher than those found by \BIZt, but are within a factor of 2. The maps of CH$_3$OH presented here are a factor of 3-20 noisier than the observations presented by \BIZt, due to the higher spectral resolution of these new maps. This can cause the difference seen in column densities. Nevertheless, the signal-to-noise ratio in these new maps is above 3$\sigma$ in the whole map for the two lines studied here, being >20$\sigma$ towards the methanol peak.

For C$^{17}$O, the column density was derived assuming optically thin emission and a constant temperature of 10\,K across the core. This temperature is based on \citetads{1999ApJ...523L.165C}, who obtained an excitation temperature of 10\,K combining the emission of C$^{17}$O (1-0) and (2-1). The hyperfine structure was fitted to check possible optical-depth effects, finding a good match with the optically thin emission assumption, as previously reported by \citetads{1999ApJ...523L.165C}.

In the case of H$_2\,^{13}$CO, HDCO and D$_2$CO, a constant temperature of 7\,K was used. This temperature comes from the modeling done for H$_2$CO (see Section \ref{h2co}), and is in agreement with that used by \citetads{2003ApJ...585L..55B}. The line D$_2$CO (2$_{1,2}$-1$_{1,1}$) is due to para species, so to obtain the total column density of D$_2$CO, the corresponding energies, degeneracy and the total partition function were used assuming an ortho-to-para ratio of 2:1. This same procedure was applied to H$_2\,^{13}$CO (2$_{1,2}$-1$_{1,1}$), which is due to ortho species, assuming and ortho-to-para ratio of 3:1. The column density map of H$_2\,^{13}$CO is used to derive a column density map of H$_2$CO assuming [$^{12}$C]/[$^{13}$C] = 77 \citepads{1994ARA&A..32..191W}.

All this information, together with the spectroscopic parameters used in the column density derivations, are summarized in Table~\ref{table_dataq}. The resulting column density maps can be found in Appendix~\ref{appendixb}. Here we only present the values at the dust and methanol peaks shown in Table~\ref{table2}. We remark that in this table the value for the column density of H$_2$CO is the one derived in Section~\ref{h2co}.    

We note that the $E_u$ and $A$ values (see Table~\ref{table_dataq}) of the different molecules are similar, and therefore, the errors associated with the assumptions of the above excitation temperatures are within the errors associated with the noise of the data. This can also be seen from the error of the column densities of CH$_2$DOH and CH$_3$OH (see Table \ref{table2}). These errors have been derived from the maximum spread of column density values found for the three different excitation temperatures and the two lines used for each species. The adopted assumption on the excitation temperature does not bias significantly our results, because said errors are $\lesssim$25\% of the column density values. Nevertheless, further observations are needed to improve in the determination of the excitation temperatures.

\begin{table*}
\centering
{\renewcommand{\arraystretch}{1.2}
\caption[Spectroscopic parameters]{Parameters used in the derivation of the column densities: excitation temperature and the corresponding partition function, the energy of the upper levels of the transitions relative to the ground states, the Einstein coefficients and the degeneracy of the upper levels. For the case of C$^{17}$O, these parameters correspond to the isolated hyperfine component. The references for these spectroscopic parameters can be found in Table~\ref{table1}.}
\label{table_dataq}
\begin{tabular}{cccccc}
\hline
\hline
Line & $T_{ex}$ (K) & $Q(T_{ex})$ & $E_u$/$k_b$ (K) & $A$ (10$^{-5}$ s$^{-1}$) & $g_u$ \\
\hline

CH$_3$OH (A$^+$) & 5, 6.5, 8 &  5.46, 10.09,  15.37 & 6.96 & 0.34 & 5 \\ 
CH$_3$OH (E$_2$) & 5, 6.5, 8 &  5.46, 10.09,  15.37 & 12.53 & 0.256 & 5 \\ 
CH$_2$DOH (2$_{0,2}$-1$_{0,1}$) & 5, 6.5, 8 & 9.60, 15.25, 22.45 & 6.40 & 0.202 & 5  \\
CH$_2$DOH (3$_{0,3}$-2$_{0,2}$) & 5, 6.5, 8 & 9.60, 15.25, 22.45 & 12.83 & 0.730 & 7 \\

H$_2\,^{13}$CO (2$_{1,2}$-1$_{1,1}$) & 7 & 8.95 & 21.72 & 4.931 & 15 \\

HDCO  (2$_{1,1}$-1$_{1,0}$)& 7 & 7.67 & 12.23 & 4.591 & 5 \\

D$_2$CO (2$_{1,2}$-1$_{1,1}$) & 7 & 15.64 & 13.37 & 2.583 & 5 \\

C$^{17}$O (1-0) & 10 & 24.37 & 5.39 & 0.0067 & 6  \\

\hline
\end{tabular}}

\end{table*}

\begin{table}
\centering
{\renewcommand{\arraystretch}{1.2}
\caption[Column densities at the center and at the methanol peak]{Column densities of the molecules observed. The value derived for H$_2$CO comes from the average between the results of two radiative transfer models, as described in Section~\ref{h2co}, and the error is the difference between both values. }
\label{table2}
\begin{tabular}{ccc}
\hline
\hline
Line & N$_{\mathrm{Dust \, peak}}$  & N$_{\mathrm{Methanol \, peak}}$  \\
 & (10$^{12}$ cm$^{-2}$) & (10$^{12}$ cm$^{-2}$)   \\
\hline
CH$_3$OH & \,\,39 $\pm$ 4 \,\,& \,\,59 $\pm$ 6 \,\,  \\ 
CH$_2$DOH & \,\,2.8 $\pm$ 0.7 & \,\,3.3 $\pm$ 0.8  \\
H$_2$CO & \,\,9 $\pm$ 2 & \,\,7 $\pm$ 3  \\
H$_2\,^{13}$CO & \,\,0.54 $\pm$ 0.35 & \,\,0.78  $\pm$ 0.35 \\
HDCO & \,\,1.30 $\pm$ 0.09 & \,\,0.96 $\pm$ 0.09  \\
D$_2$CO & \,\,1.5 $\pm$ 0.3 & \,\,1.1 $\pm$ 0.3  \\
C$^{17}$O & 680 $\pm$ 60 & 810 $\pm$ 60  \\

\hline

\end{tabular}}

\end{table}

\subsection{Deuterium fraction}
\subsubsection{CH$_2$DOH/CH$_3$OH}
\begin{figure}
\resizebox{\hsize}{!}{\includegraphics{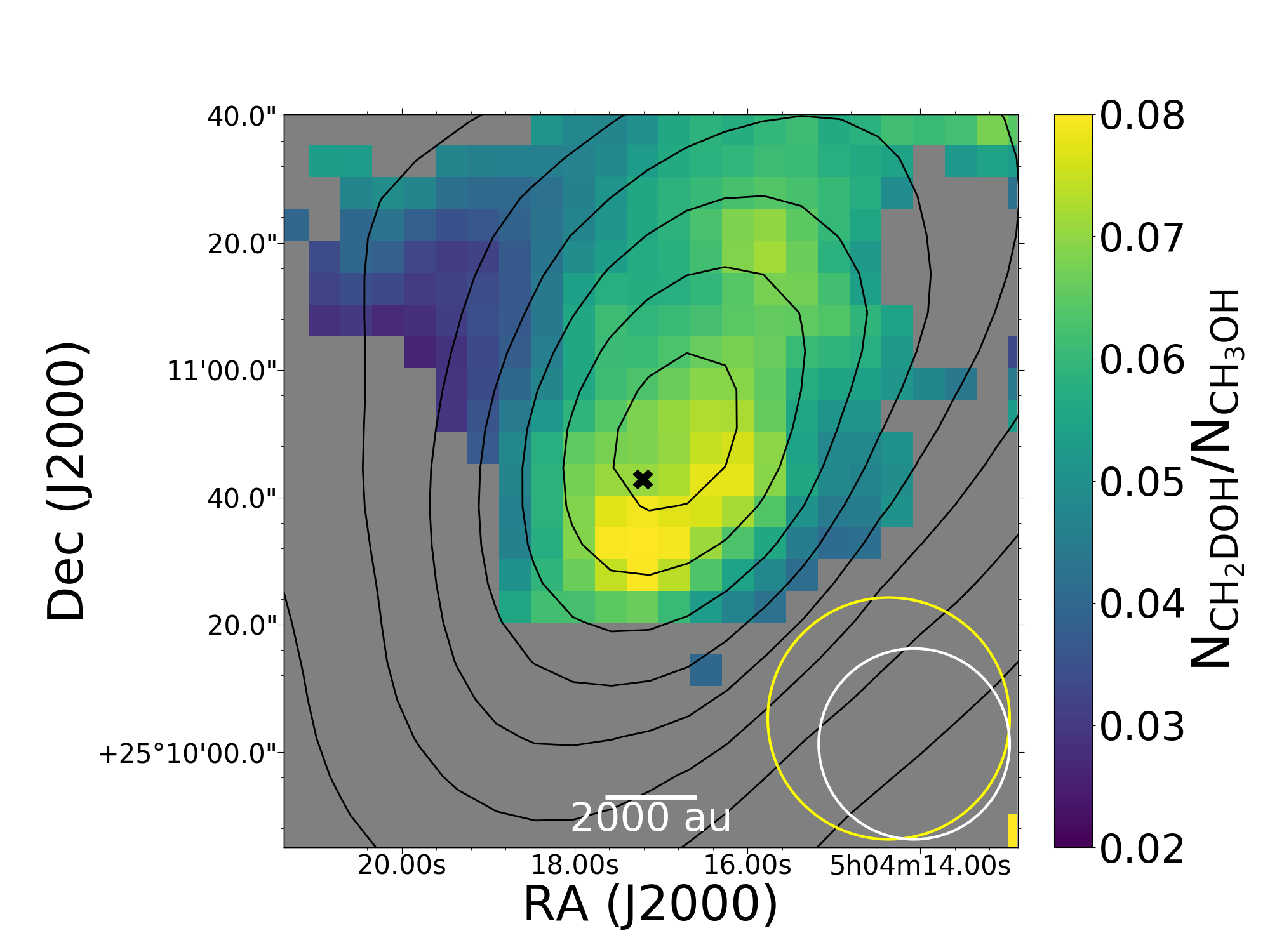}} 
\caption[Deuterium fraction map of methanol]{Deuterium fraction map of methanol using the column density maps of CH$_2$DOH and CH$_3$OH derived as described in Section \ref{cdens}. The black contours represent increasing 10\% steps of the peak of the Herschel N(H$_2$) map, presented by \SPEt. The HPBWs are shown in the bottom right corners of the figures, in yellow for \textit{Herschel}/SPIRE and in white for the 30m telescope. The black cross marks the dust continuum peak, and only values above the 3$\sigma$ detection level are considered. } 
\label{deuteration_maps}
\end{figure}

The deuterium fraction map of methanol was obtained by taking the ratio between the column density of CH$_2$DOH and that of CH$_3$OH. As seen in Fig.~\ref{deuteration_maps}, it shows a non symmetric distribution peaking toward the south but close to the dust peak, especially when the beam size is taken into account. The [CH$_2$DOH]/[CH$_3$OH] peak value, $\sim$0.08$\pm$0.02, is consistent with that found by \BIZt\, toward the core center.

\subsubsection{H$_2$CO, HDCO and D$_2$CO}

\begin{figure*}
\resizebox{\hsize}{!}{\includegraphics{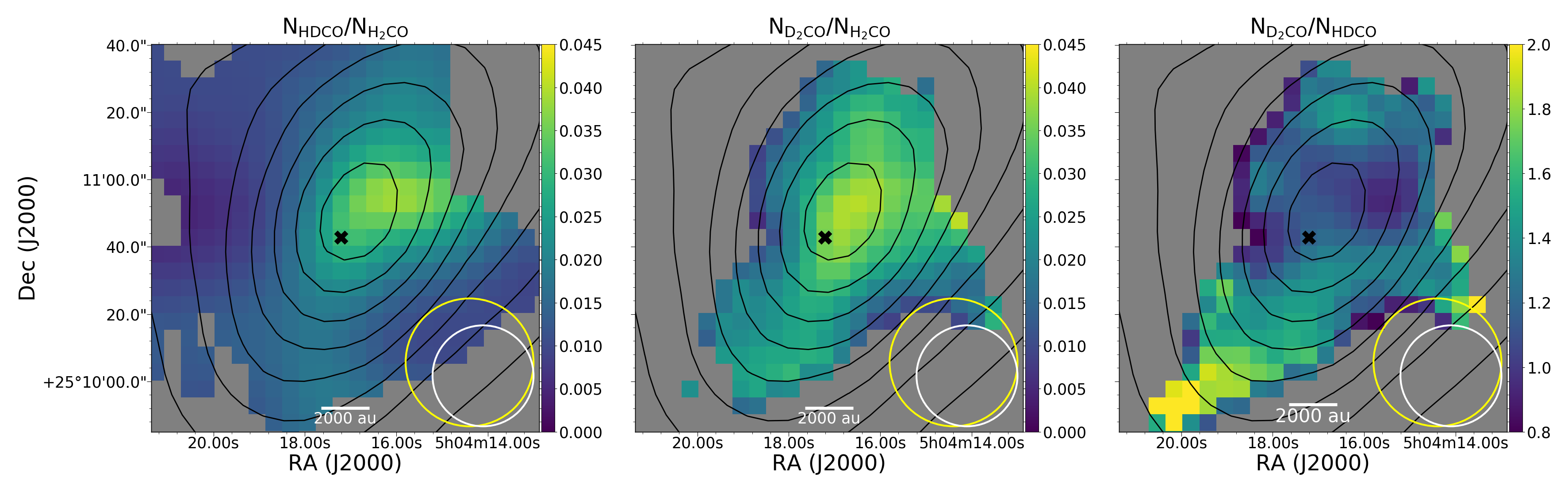}} 
\caption[Deuterium fraction maps of formaldehyde]{ \textit{Left panel}, [HDCO]/[H$_2$CO] map using the column density maps of HDCO and H$_2$CO derived as described in Section \ref{cdens}. \textit{Middle panel}, map of the ratio [D$_2$CO]/[H$_2$CO], using the column density maps of D$_2$CO and H$_2$CO derived as described in Section \ref{cdens}. \textit{Right panel,} map of the ratio [D$_2$CO]/[HDCO], using the column density maps of HDCO and D$_2$CO derived as described in Section \ref{cdens}. In all panels, the black contours represent increasing 10\% steps of the peak of the Herschel N(H$_2$) map, presented by \SPEt. The HPBWs are shown in the bottom right corners of the figures, in yellow for \textit{Herschel}/SPIRE and in white for the 30m telescope. The black cross marks the dust continuum peak, and only values above the 3$\sigma$ detection level are considered. } 
\label{h213co-hdco-d2co}
\end{figure*}

The deuterium fraction maps of formaldehyde are shown in Fig.~\ref{h213co-hdco-d2co}. They are derived using the column density maps of HDCO, D$_2$CO and H$_2$CO. The H$_2$CO, as commented previously, is derived from the map of the optically thin emission of H$_2\,^{13}$CO. The maps show that the deuteration of formaldehyde also occurs in the central regions of the core. However, this needs more sensitive observations because of the large uncertainties, which can be seen in Fig.~\ref{deut_h2co_error}. Thus, no information regarding the distribution can be obtained. As reference, the values at the dust peak are: [HDCO]/[H$_2$CO]$\simeq$0.03$\pm$0.02 and [D$_2$CO]/[H$_2$CO]$\simeq$0.04$\pm$0.03.

The map of the [D$_2$CO]/[HDCO] ratio shows a very high level of deuteration, with a mean value of $\sim$1.2$\pm$0.3. As previously mentioned,  any variation of the deuteration across the map is not significant due to the high errors in its determination (see Fig.~\ref{deut_h2co_error}).

\section{Comparison with models}\label{models2}
In this section, observational and theoretical results are compared. Here we present two different chemical models applied to L1544, and use them to compare the modeled column densities with the observed ones and to estimate the H$_2$CO column density. The two chemical models used are:
\begin{itemize}
\item {\bf{S16:}}
In this case the abundances of $\rm C^{17}O$, $\rm CH_3OH$, $\rm CH_2DOH$, $\rm H_2CO$, HDCO, and $\rm D_2CO$ were calculated using the chemical model presented in \citetads{Sipila15a,Sipila15b}, which includes extensive descriptions of deuterium and spin-state chemistry. The modeling process was similar to that discussed in \citetads{Sipila16a}, i.e. we divided the L1544 core model presented by \citetads{2010MNRAS.402.1625K} (see also \citealt{Keto14}) into concentric shells, calculated the chemistry separately in each shell, and extracted the abundance gradients of the various species at several time steps. The model uses a bulk ice model, and it includes a fixed reactive desorption efficiency. The results shown in Figs. \ref{h2co_anton_scaled}, \ref{model_ch3oh},\ref{model_hdco}, \ref{deut_model_ch3oh} and \ref{deut_model_hdco} correspond to the time when the CO column density is comparable to the observed value in L1544 \citepads{1999ApJ...523L.165C}.  

\item {\bf{V17:}}
The model of the pre-stellar core L1544 by \citetads{2017ApJ...842...33V} utilized static 1D distributions of gas density, gas and dust temperatures from \citetads{Keto14}. The model includes time-dependent description of gas and grain chemistry based on \citetads{2013ApJ...769...34V} with several major updates concerning new gas-phase reactions important for the formation of complex organic molecules in the cold gas, as well as the detailed treatment of the efficiency of reactive desorption based on experimental works by \citetads{2016A&A...585A..24M}. Treatment of chemistry on interstellar grains is based on a multilayer approach to the structure of icy mantles, which allows to discriminate between the reactive surface of ice and more chemically inert ice bulk. Since the temperature in L1544 is $\sim$10\,K, the only important source of mobility of species on ice surfaces is quantum tunneling, which is enabled for H and H$_2$. 
\end{itemize}

Both H$_2$CO and CH$_3$OH in the models are produced on the surface in the hydrogenation sequence of CO:
\begin{equation}
\phantom{AAA}
\mathrm{CO} \rightarrow \mathrm{HCO} \rightarrow \mathrm{H_2CO} \rightarrow \mathrm{CH_2OH} \rightarrow \mathrm{CH_3OH} \notag
\end{equation}
However, while surface route is the only efficient way of formation for CH$_3$OH, H$_2$CO is also formed efficiently in the gas-phase reaction
\begin{equation}
\phantom{AAA}
\mathrm{O} + \mathrm{CH_3} \rightarrow \mathrm{H_2CO} + \mathrm{H}. \notag
\end{equation} 

In summary, both models use the same physical cloud structure, do not follow the evolution of the density and temperature of the cloud with time (here called static models), and adopt the same chemical reactions. However, the S16 model uses a bulk ice model with a fixed reactive desorption efficiency, it includes deuterated species, allows diffusion of light species via thermal hopping, and starts its chemistry from atomic initial abundances. Meanwhile, V17 does not include deuterated isotopologues, but uses a multilayer approach for the ices with an updated treatment of the reactive desorption efficiency, and also allows hydrogen atoms and molecules to quantum tunnel on the grain surfaces, which facilitates the formation of methanol and other complex organic molecules (COMs). Moreover, this model assumes the evolved chemistry of diffuse clouds as initial conditions. However, no matter what initial conditions we use (i.e. if we start our calculation with atomic conditions, typical of diffuse clouds, or if we start with conditions typical of molecular clouds), the final modeled results for our pre-stellar core chemical composition are the same. Therefore, different initial conditions do not play a role in the different results from the models concerning methanol; the differences here can solely be ascribed to the different diffusion mechanisms of H on the surface, with V17 allowing quantum tunneling and S16 allowing the significantly slower thermal hopping. The comparison between models is important to check the dominant processes in the formation of COMs.  

\subsection{Deriving the column density of H$_2$CO}\label{h2co}
As already stated, H$_2$CO presents signatures of self-absorption, so one cannot apply the Equation \eqref{eqcolumn} and the assumption of optical thinness to this molecule. Thus, we intended to model the emission at the center of the core via radiative transfer modeling using MOLLIE \citepads{2010ApJ...716.1315K}, the physical model for L1544 from \citetads{2015MNRAS.446.3731K}, and the abundances profiles obtained from S16 and V17. The resulting column densities from MOLLIE are smoothed with a 30\arcsec\, beam to match the resolution of the observations. This procedure was only used for comparing the column density derived from modeling with that observed, derived from the H$_2\,^{13}$CO line.

\begin{figure*}[h]
\resizebox{\hsize}{!}{\includegraphics{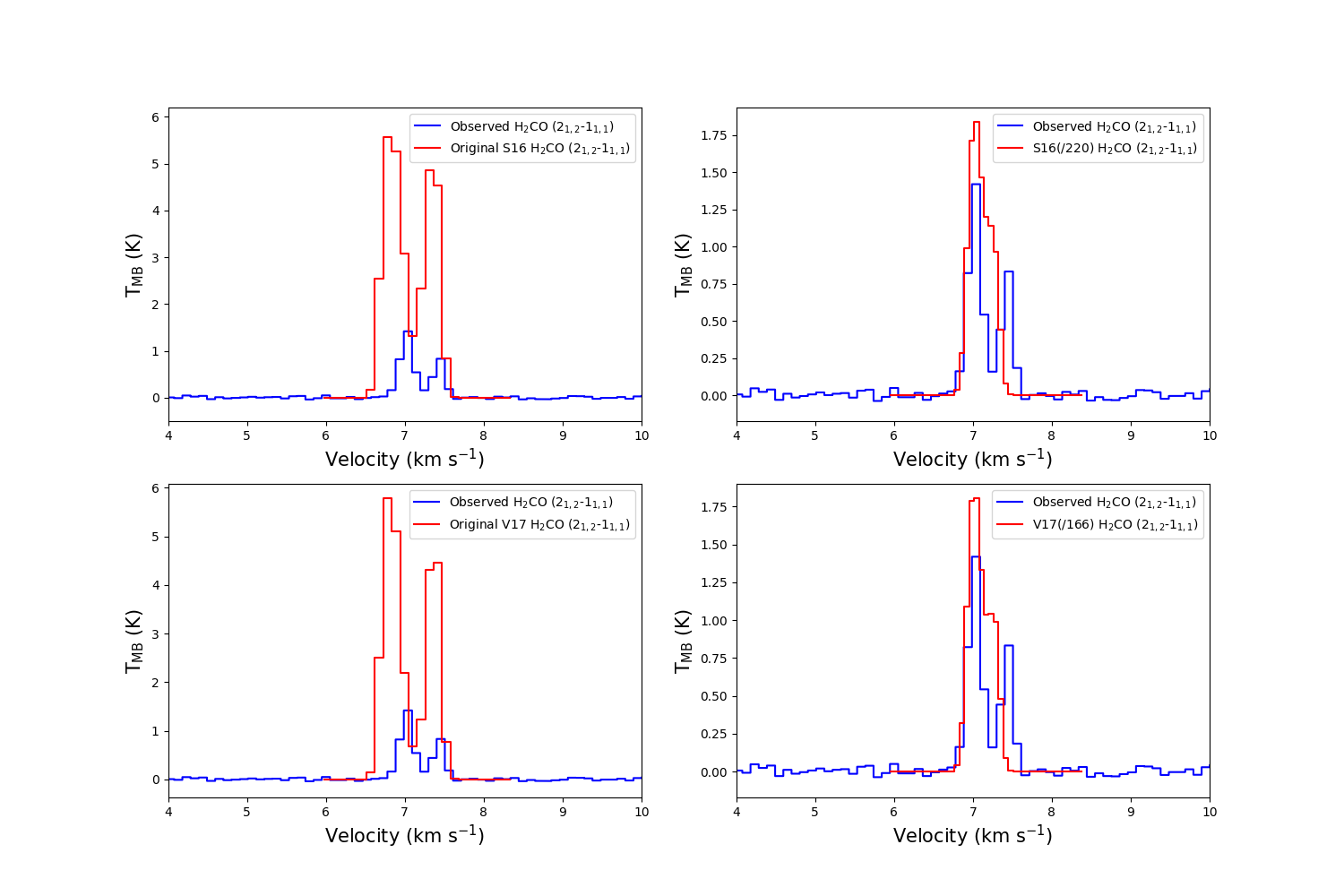}} 
\caption[Comparison between the modeled and observed emission of H$_2$CO]{Comparison of the emission modeled by MOLLIE (in red, see Section \ref{h2co}) and the observed emission (in blue) of the H$_2$CO transition observed at the center of the core. \textit{Top panels:} results using S16 abundance profile (\textit{left panel}), and the same scaled by a factor of 220 (\textit{right panel}). \textit{Bottom panels:} results using V17 abundance profile (\textit{left panel}), and the same scaled by a factor of 166 (\textit{right panel}). The original abundance profiles overestimate the emission of the line by a factor of $\sim$3, and the scaled abundance profiles do not reproduce the shape of the line.} 
\label{h2co_anton}
\label{h2co_anton_scaled}

\end{figure*}

We found that both abundance profiles overestimated the emission by a factor of $\sim$3 (see Fig.~\ref{h2co_anton}). Thus, we scaled the abundance profiles to reproduce the observed peak emission. The scaling factor applied to the abundance profile of V17 was 166, while the abundance profile from S16 needed to be reduced by a factor 220. The fact that the abundances had to be reduced by such  large factors is due to the high optical depth of the line. The resulting modeled column densities can be found in Table \ref{table2}. The excitation temperature obtained in both cases was of $\sim$7 K in the central 0.1 pc, and this value has been adopted as the temperature used for the derivation of the HDCO, D$_2$CO and H$_2\,^{13}$CO column densities. 
At the center of the core, the column density of H$_2$CO derived from H$_2\,^{13}$CO is a factor of 5 higher than the column density derived here using MOLLIE. This difference is below 2$\sigma$ (being $\sigma$ the error in column density derived observationally), once the uncertainties are taken into account.

The abundance reduction produces a change in the line profile, which now does not show the observed double peaked asymmetry any longer (see Fig.~\ref{h2co_anton_scaled}). One explanation is that the models do not take into account the presence of a lower density molecular material surrounding the core, where H$_2$CO can be present and participate in the line absorption. In fact, the results from the H$_2\,^{13}$CO line suggest an optical depth of $\tau\sim$\,1.4 for the H$_2$CO line, while the optical depth predicted by MOLLIE of the H2CO line is $\tau\sim$\,0.3. 
The ones presented here are therefore only approximate estimates of the column density, and although they are not far from the observations, more detailed work (i.e. larger scale mapping of H$_2$CO) is needed to fully understand from which layer of the cloud the formaldehyde emission is coming from.

\subsection{Modeled vs. observed column densities}

In this section, the column densities observed are compared with the results from the V17 and S16 models. The V17 model does not include deuterated species so, in this case, the modeled column densities of deuterated species were derived using the column density of their corresponding isotopologue and the deuteration ratio from S16. 

\begin{figure*}
\centering
\includegraphics[width=1\textwidth]{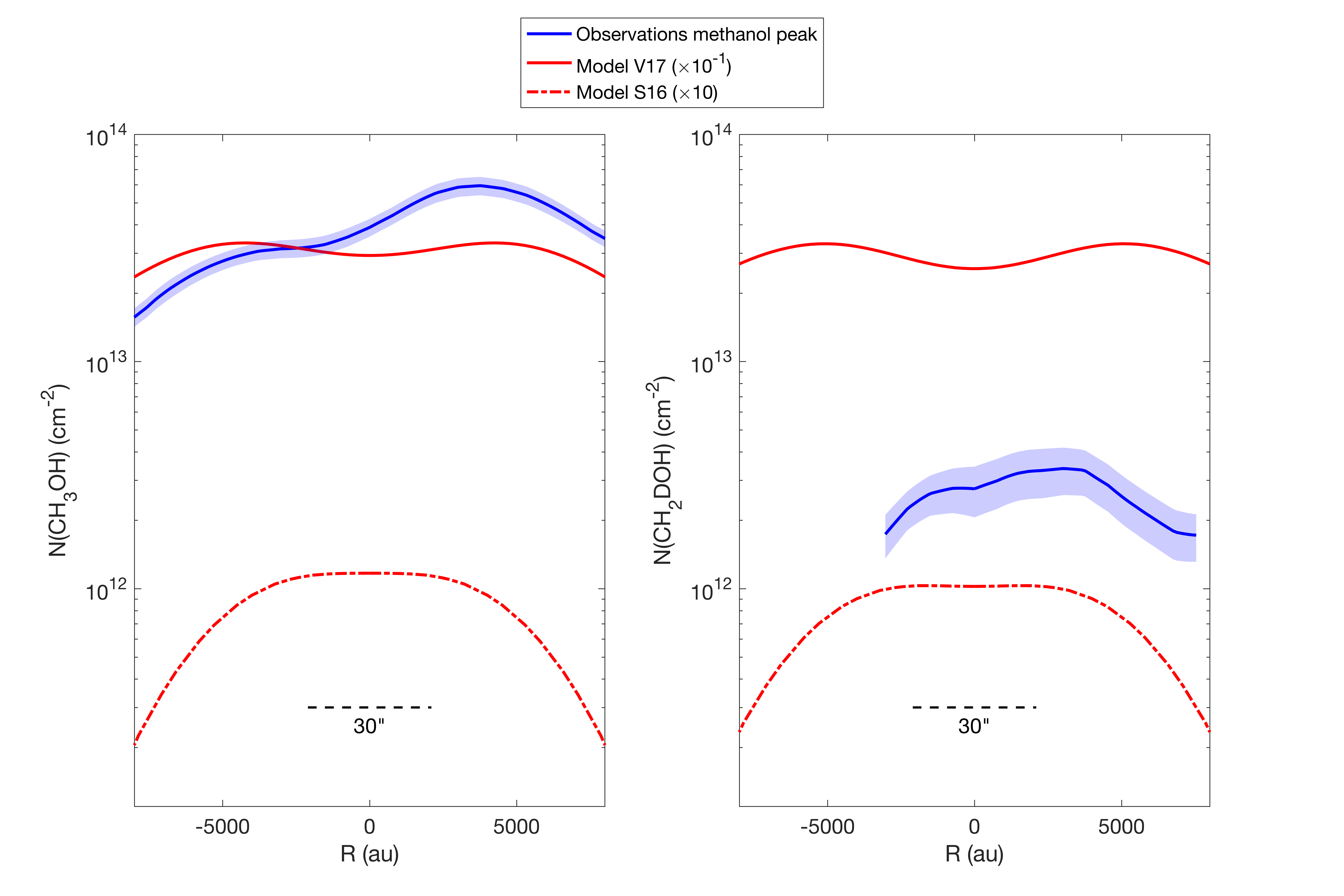}
\caption[Comparison of the observed and the modeled column densities of methanol and deuterated methanol]{Comparison of the observed column densities (blue line) with the modeled column densities by V17 (red line) and S16 (red dashed line) of CH$_3$OH (\textit{left panel}) and CH$_2$DOH (\textit{right panel}). The observed column density profiles are taken along the white dashed line shown in Fig.~\ref{moment0_ch3oh-ch2doh}, being R=0 au the dust peak, R>0 au the direction from dust peak to methanol peak, and R<0 au is the direction from the dust peak toward the south--west. The shaded blue regions indicate the error bars of the column densities. The resolution is 30\arcsec, shown at the bottom of the figure.  Note the scaling factor applied to the models.} 
\label{model_ch3oh}
\end{figure*}

\begin{figure*}
\centering
\includegraphics[width=1\textwidth]{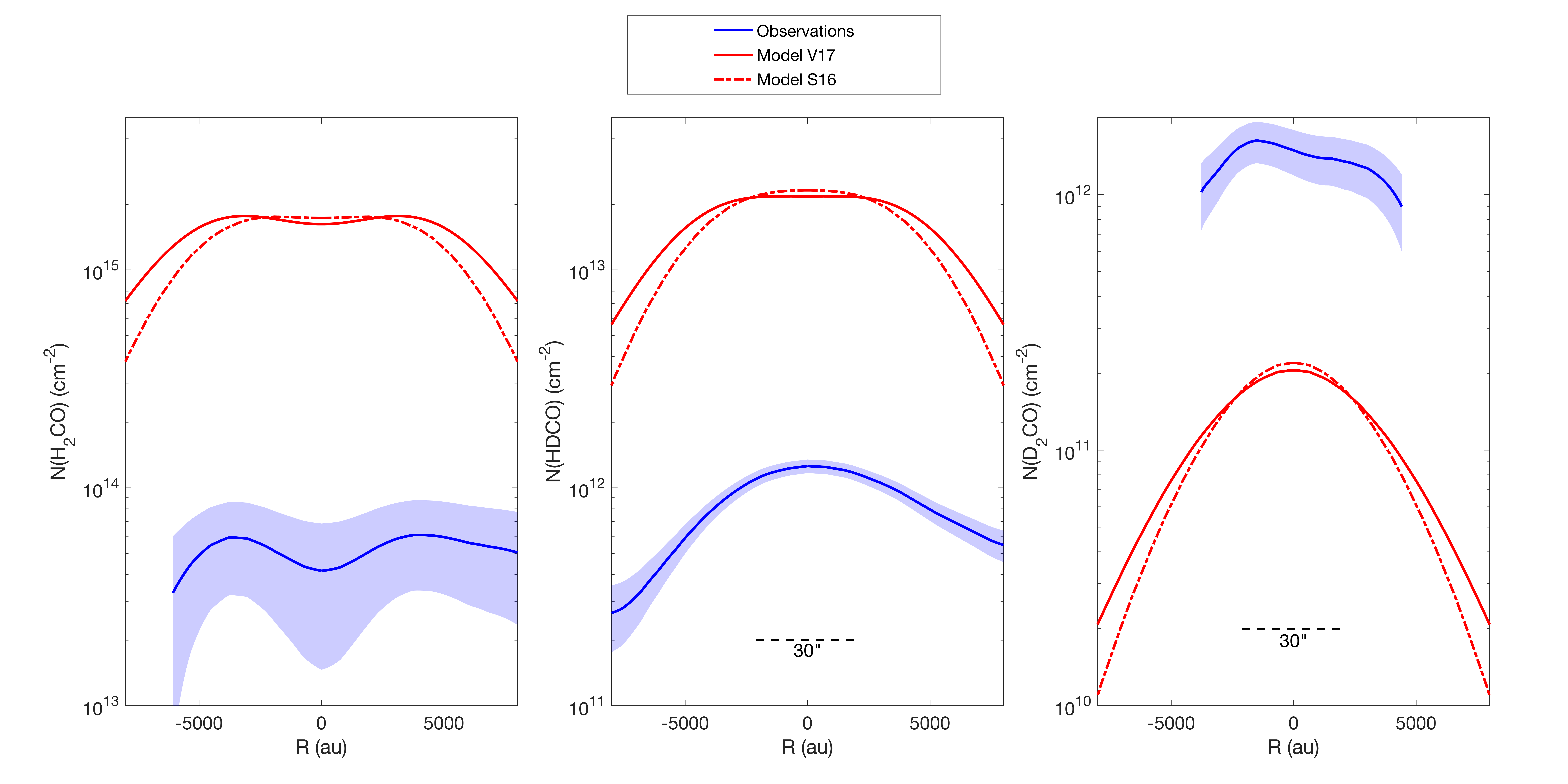}
\caption[Comparison of the observed and the modeled column densities of formaldehyde and its deuterated species]{Comparison of the observed column densities (blue line) with the modeled column densities by V17 (red line) and S16 (red dashed line) of H$_2$CO (\textit{left panel}), HDCO (\textit{middle panel}) and D$_2$CO (\textit{right panel}). The observed column density profiles are taken along the white dashed line shown in Fig.~\ref{moment0_ch3oh-ch2doh}, being R=0 au the dust peak, R>0 au the direction from dust peak to methanol peak, and R<0 au is the direction from the dust peak toward the south--west. The shaded blue regions indicate the error bars of the column densities. The resolution is 30\arcsec, shown at the bottom of the figure.} 
\label{model_hdco}
\end{figure*}

The results can be found in Fig.~\ref{model_ch3oh} for methanol and deuterated methanol and in Fig.~\ref{model_hdco} for single and doubly deuterated formaldehyde (we note the scaling factor applied to the models in some figures, indicated in the legend). The observed column densities are those corresponding to a cut done in the maps, which follows the white dashed line showed in  Fig.~\ref{moment0_ch3oh-ch2doh}. This cut is the same as the one studied by \SPEt\, who considered the variation of the methanol column density from the methanol peak to the dust peak. We point out that the same conclusions are reached if the cut orientation is changed, as the methanol column density changes by a factor of a few around the asymmetric ring, while the difference with the model prediction is larger than one order of magnitude. 

Methanol and deuterated methanol are overestimated in the model from V17, and underestimated by S16. The overproduction of CH$_3$OH in V17, which was already noticed by the authors, can be due to an overestimation of the efficiency of reactive desorption. Moreover, as explained in V17, this model does not efficiently form CO$_2$ because of the activation energy of the CO+OH reaction \citepads{2001MNRAS.324.1054R}, which does not proceed at T<20\,K \citepads{2017ApJ...842...33V}, so in warmer environments, one would expect to have a significant fraction of CO producing CO$_2$ instead of methanol. CO$_2$ is also abundant and ubiquitous in quiescent dense clouds \citepads{1998ApJ...498L.159W,2015ARA&A..53..541B}; however, in these cold regions, CO$_2$ is expected to form via energetic processing of water ice mantles onto carbonaceous grains \citepads[see e.g.]{2004ApJ...615.1073M,2006ApJ...643..923M}, and this process has not been included in any gas-grain chemical model yet. In any case, we note that a factor of 10 difference between model and observations can be considered to be in fair agreement, due to the large uncertainties of gas-grain models, which can go up to one order of magnitude \citepads{2004AstL...30..566V, 2008ApJ...672..629V}. The underproduction in S16 is most likely caused by the fact that hydrogen atoms are not allowed to quantum tunnel across the grain surface, thus significantly reducing surface hydrogenation and the consequent production of CH$_3$OH.

\begin{figure}
\resizebox{\hsize}{!}{\includegraphics{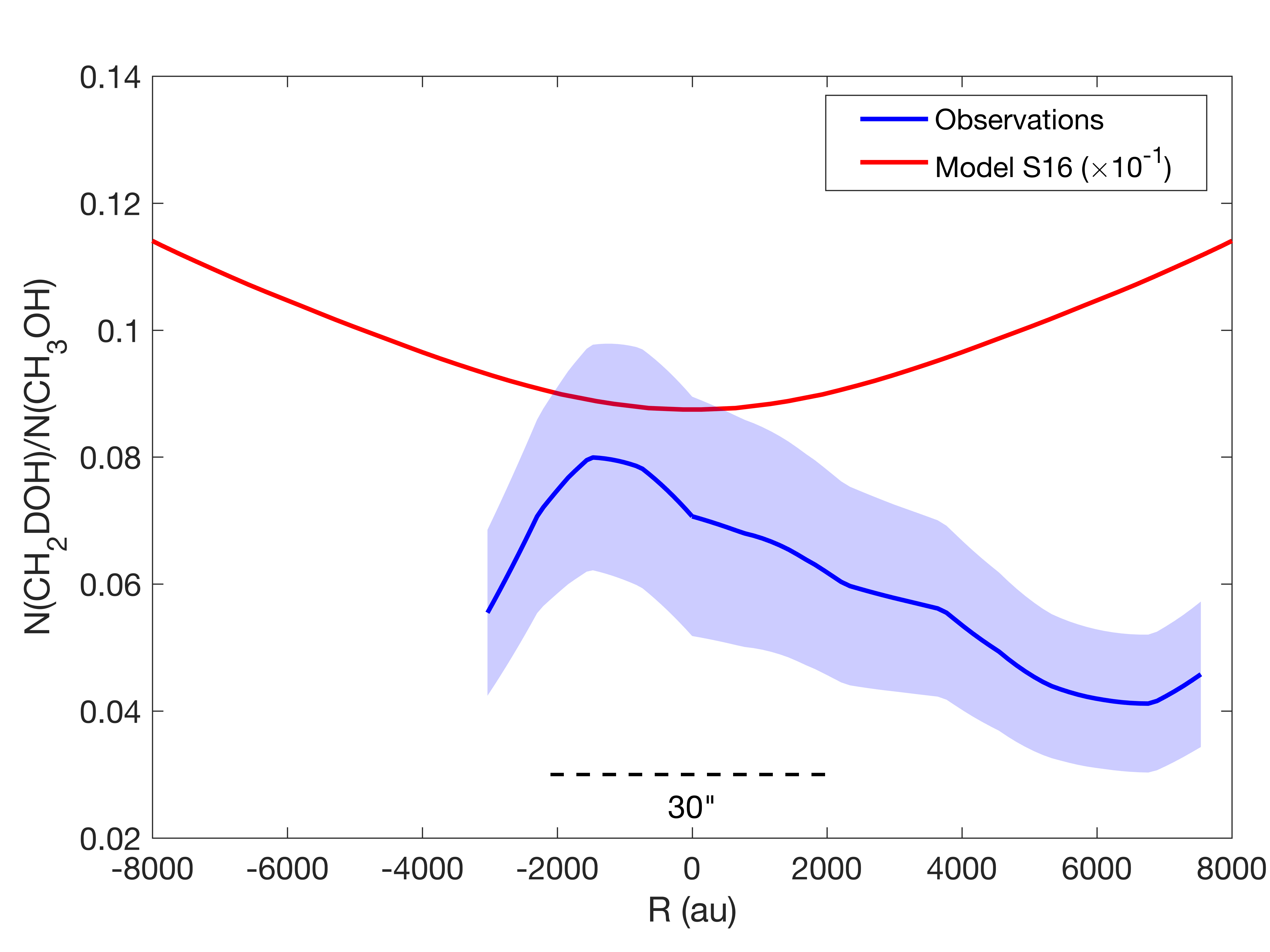}} 
\caption[Comparison between the observed and the modeled deuterium fraction of methanol]{Comparison of the observed methanol deuterium fraction (blue line) with the modeled deuterium fraction by S16 (red line).  The observed deuteration profile is taken along the white dashed line in Fig.~\ref{moment0_ch3oh-ch2doh}, being R=0 au the dust peak, R>0 au the direction from dust peak to methanol peak, and R<0 au is the direction from the dust peak toward the south--west. The shaded blue region shows the error bars in the deuterium fractions. The resolution is 30\arcsec, shown at the bottom of the figure.  Note the scaling factor applied to the model.}
\label{deut_model_ch3oh}
\end{figure}

The measured column density of methanol is closer to the V17 predictions, whereas deuterated methanol is closer to S16. One has to take into account that for the models, the deuterium fraction was taken from S16, which overestimates the deuteration of methanol by a factor of 10 (see Fig.~\ref{deut_model_ch3oh}). If the observed deuterium fraction ($\sim$0.1) is applied to the models, the difference between observations and both models is similar.

\begin{figure*}
\resizebox{\hsize}{!}{\includegraphics{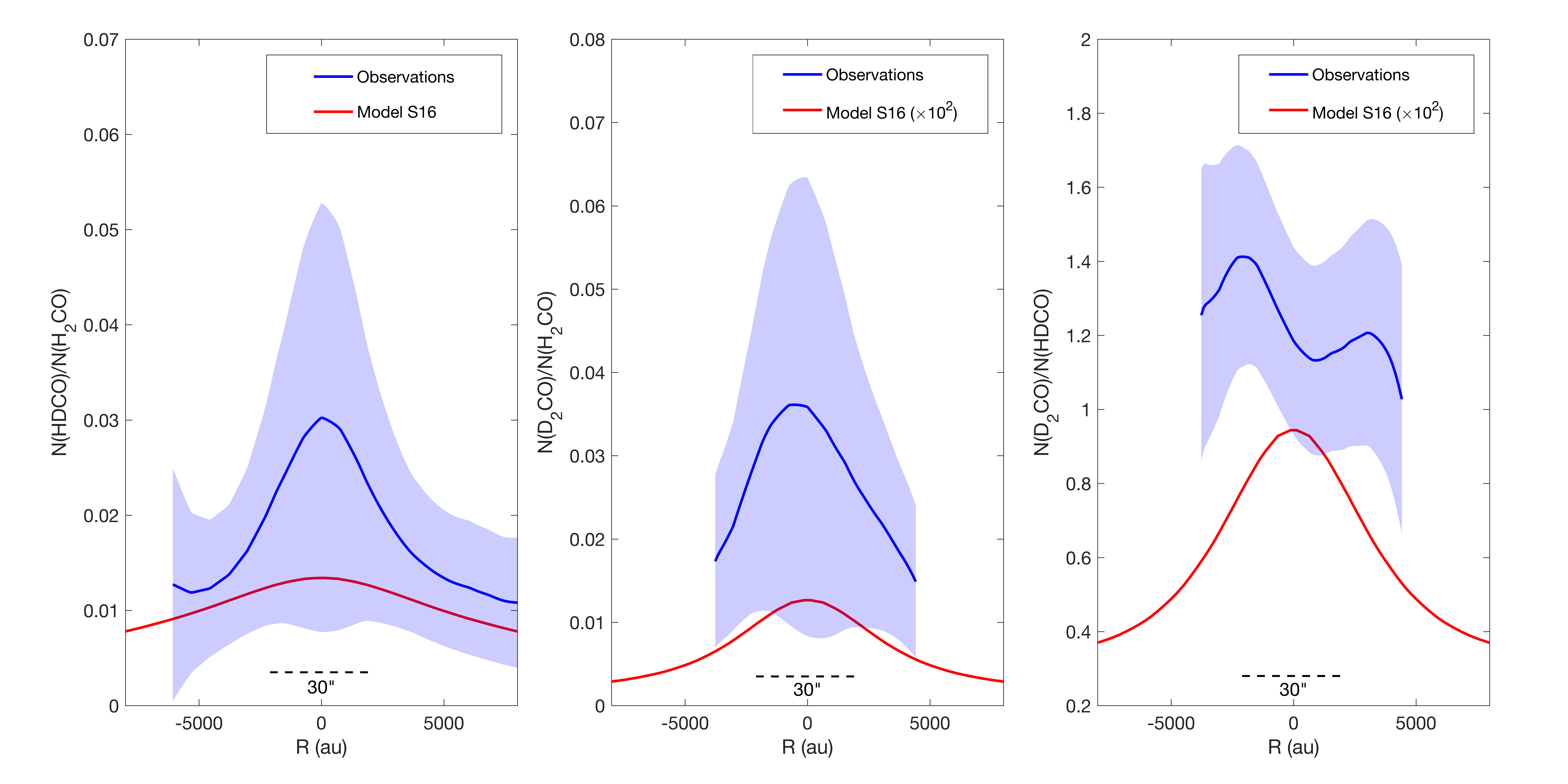}} 
\caption[Comparison between the observed and the modeled deuterium fractions of formaldehyde and deuterated formaldehyde]{Comparison of the observed [HDCO]/[H$_2$CO] (\textit{left panel}), [D$_2$CO]/[H$_2$CO] (\textit{middle panel}) and [D$_2$CO]/[HDCO] (\textit{right panel}) ratios (blue line) with the modeled deuterium fractions by S16 (red line). The observed deuteration profile is taken along the white dashed line in Fig.~\ref{moment0_ch3oh-ch2doh}, being R=0 au the dust peak, R>0 au the direction from dust peak to methanol peak, and R<0 au is the direction from the dust peak toward the south--west. The shaded blue regions indicate the error bars in the deuterium fractions. The resolution is 30\arcsec, shown at the bottom of the figure.  Note the scaling factor applied to the models. }
\label{deut_model_hdco}
\end{figure*}

In the case of formaldehyde, both models overproduce H$_2$CO. For its deuteration, S16 model predicts [HDCO]/[H$_2$CO] $\sim$0.013, and [D$_2$CO]/[H$_2$CO] a factor of 100 smaller (see Fig.~\ref{deut_model_hdco}). However, observationally, we found a value of $\sim$0.03$\pm$0.02 for [HDCO]/[H$_2$CO] and $\sim$0.04$\pm$0.03 for [D$_2$CO]/[H$_2$CO]. This means that while [HDCO]/[H$_2$CO] is well reproduced by the model, there is a difference of a factor of 100 between the observed [D$_2$CO]/[H$_2$CO] and the modeled ratio. Applying this observed factor to the modeled abundances of H$_2$CO would imply an overproduction of all the formaldehyde family (see Fig.~\ref{model_hdco}). 

Our results show that the production of these molecules is still uncertain in this core, and more work has to be done to better constrain the chemistry in pre-stellar cores.

\section{Discussion}\label{discussion2}

\subsection{Distribution}

\begin{figure}
\resizebox{\hsize}{!}{\includegraphics{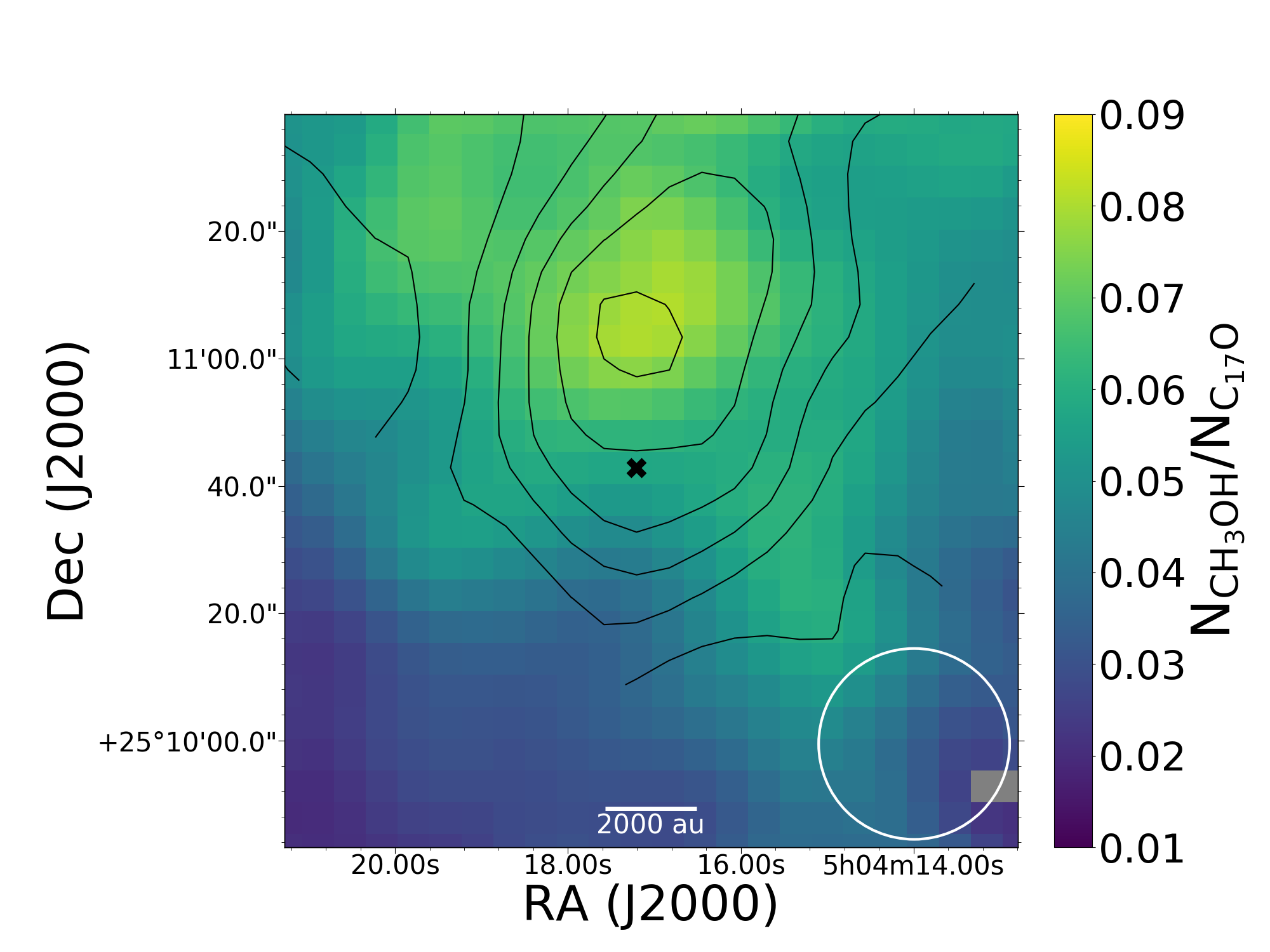}} 
\caption[Map of the N(CH$_3$OH)/N(C$^{17}$O) ratio]{Ratio of the column densities of methanol and C$^{17}$O. The black contours represent 10\% steps with respect to the deuterated methanol column density peak. The black cross marks the dust peak. The HPBW is shown in the bottom right corner. } 
\label{ch3oh-c17o-ch2doh}
\end{figure}

The fact that deuterated methanol peaks closer to the core center compared to methanol can be simply explained by the larger D/H abundance, so that D atoms compete with H atoms in saturating solid CO. This central region is in fact characterized by large amount of CO freeze-out \citepads{1999ApJ...523L.165C} and consequently large deuterium fractions \citepads{2002ApJ...565..344C, 2003A&A...403L..37C, 2007A&A...470..221C}, as predicted by theory \citepads{1984ApJ...287L..47D, 2004A&A...418.1035W}. This scenario is supported by the fact that deuterated methanol peaks exactly where the N(CH$_3$OH)/N(CO) ratio is higher (see Fig.~\ref{ch3oh-c17o-ch2doh}), and the deuterium fraction is enhanced where the CO is more depleted (see Fig. \ref {c17o_columndens}). 

For formaldehyde, we find that it follows a slightly different distribution compared to CH$_3$OH, interestingly peaking at the same place as C$^{17}$O and showing a secondary peak toward the south--west part of the core, where the emission of C$^{17}$O also shows a decrease. This can be explained by the fact that, unlike CH$_3$OH, H$_2$CO can also form via gas phase routes involving hydrocarbons (e.g. CH$_2$ and CH$_3$; \citeads{2017iace.book.....Y}), which are expected to be abundant toward the southern part of the core, where the extinction abruptly drops to relatively low values and where carbon chains are in fact abundant \citepads{2016A&A...592L..11S, 1999ApJ...518L..41O}.

The deuterated species of H$_2$CO both peak at the center of the core, which is not the case for deuterated methanol, as seen in Fig.~\ref{moment0_hdco-d2co}. Their distributions follow that of other deuterated species as N$_2$D$+$ and NH$_2$D \citepads{2002ApJ...565..344C, 2007A&A...470..221C}. 

Interferometric observations are needed to make more detailed conclusions on possible differences in the distributions of the deuterated forms of methanol and formaldehyde, as the  resolution and sensitivity of these observations do not allow us to discuss them in locations further than where their emission is enhanced.

\subsection{Deuteration}

Previous results for the deuterium fractions measured in L1544 show a wide range of values: [N$_2$D$^+$]/[N$_2$H$^+$] = 0.2, [NH$_2$D]/[NH$_3$] = 0.5, [DCO$^+$]/[HCO$^+$]=0.04 and [c-C$_3$HD]/[c-C$_3$H$_2$] =0.12-0.17 \citepads{2002ApJ...565..344C, 2007A&A...470..221C, 2013ApJ...769L..19S}. 
The CH$_2$DOH/CH$_3$OH column density ratio never reaches values larger than 0.1 (with a peak value of $\sim$0.08$\pm$0.02), unlike [N$_2$D$^+$]/[N$_2$H$^+$] and [NH$_2$D]/[NH$_3$] \citepads{2005ApJ...619..379C, 2007A&A...470..221C}. This suggests that deuterated methanol is probably tracing a more external layer than N$_2$D$^+$ and NH$_3$. This is analogous to the conclusion reached by \citetads{2002ApJ...565..344C} to explain the [DCO$^+$]/[HCO$^+$] value, since both DCO$^+$ and HCO$^+$ require gas-phase CO to form and CO is mainly frozen onto dust grains toward the core center, as already mentioned. The larger deuteration measured in N$_2$H$^+$ and NH$_3$ could also reflect the fact that these two molecules require larger times to form compared to CO, because of the slow formation rate of the parent molecule N$_2$. Thus, molecular nitrogen becomes abundant only toward the central regions of the core, where CO is mostly frozen and where H$_3\,^+$ is highly deuterated. As CO and related molecules (such as CH$_3$OH and its deuterated forms) are mainly in the solid phase in these central regions \citepads{1999ApJ...523L.165C}, deuterated methanol can only trace the zones surrounding the dust peak, where the D/H ratio is expected to be lower compared to the one deduced by the above mentioned N-bearing species.

The deuterium fraction of methanol shows lower values than those found towards more evolved Class 0 objects. This difference with more evolved objects can be due to the presence of a major reservoir of CH$_2$DOH in the ices in pre-stellar cores, which is released to the gas phase during the Class 0 phase \citepads[e.g.]{2002A&A...393L..49P, 2004A&A...416..159P}. However, this scenario is still a matter of discussion due to recent results from high resolution observations \citepads[e.g. ]{2017A&A...606L...7B, 2018A&A...610A..54P}, which show lower deuterium fractions for Class 0 objects. 

Similarly, we find that [HDCO]/[H$_2$CO]$\simeq$0.03$\pm$0.02 and [D$_2$CO]/[H$_2$CO]$\simeq$0.04$\pm$0.03 at the dust peak, which is consistent with previous values reported in pre-stellar cores and Class 0 objects \citepads{2003ApJ...585L..55B, 2006A&A...453..949P}. The value for [D$_2$CO]/[H$_2$CO] is also consistent with that found by \citetads{2003ApJ...585L..55B}. One thing to note is that, as in \citetads{2011A&A...527A..39B}, we find a higher abundance of D$_2$CO than HDCO. \citetads{2011A&A...527A..39B} claimed that this can only be explained if grain chemistry is ongoing, although their results are applied to the region of Ophiuchus, and the chemistry can differ from cloud to cloud. However, these molecules can suffer from depletion, which can be the source of discrepancy in the deuterium fractionations between different species, more importantly when compared  with those that suffer less from depletion, like N$_2$D$^+$ and NH$_3$.

As seen in Section \ref{models2}, the model predictions for the deuteration of methanol in L1544 is higher than those observed, while the opposite happens with D$_2$CO. The deuteration of formaldehyde is close to that of methanol if the uncertainties are taken into account. Thus, more theoretical work needs to be done to fully understand deuteration processes in pre-stellar cores.

\section{Conclusions}\label{conclusions2}

We have presented our maps of methanol, formaldehyde and their deuterated species toward the well known pre-stellar core L1544. These two molecules can help us to gain understanding about the chemical processes taking place on the dust grain surfaces and the formation of more complex organic molecules, as well as the deuterium history in the process of star formation. 

The highest level of deuteration of methanol occurs close to the dust peak, reaching [CH$_2$DOH]/[CH$_3$OH] = 0.8. This indicates that a more external layer is traced by CH$_2$DOH compared to N$_2$H$^+$ \citepads{2002ApJ...565..331C} and NH$_3$ \citepads{2007A&A...470..221C}. CH$_2$DOH also shows a higher abundance at a distance of $\sim$3000 au from the dust peak, exactly where more methanol is present in the gas phase with respect to CO. This suggests that deuterated methanol is formed and released the same way CH$_3$OH is.

HDCO and D$_2$CO, however, peak toward the center of the core, and present a high deuterium fraction, only found previously in $\rho$ Oph A \citepads{2011A&A...527A..39B}.  Interestingly, H$_2$CO shows a ring like structure, depleting towards the center, and showing two maxima, one coinciding with C$^{17}$O, and a secondary peak toward the South-West, unlike methanol, which coincides with a region where C$^{17}$O shows less emission. This suggests that gas phase production via reactions involving hydrocarbons efficiently takes place in regions where C is not completely locked in CO, based on the conclusions of \SPEt.

Finally, we compared two different chemical models with our observational results, and did not find an agreement. On the one hand, the model from \citetads{2017ApJ...842...33V} is overproducing methanol and formaldehyde, probably caused by the overestimation of the efficiency of reactive desorption. On the other hand, the model from \citetads{Sipila15a,Sipila15b} does not produce enough methanol because of the slow diffusion of hydrogen atoms on the surfaces (which are not allowed to quantum tunnel) and it overproduces formaldehyde. Both models are static and the inclusion of dynamical evolution could change the results, although not by orders of magnitude, needed to reconcile models with observations. Our results suggest that quantum tunnelling for H diffusion on icy dust mantles should be considered, while reactive desorption still needs more detailed experimental work. 

Higher sensitivity and angular resolution observations of the lines presented here are needed, together with a parameter-space exploration within current chemical models and laboratory work, to shed light on important chemical processes happening at the dawn of star formation.

\begin{acknowledgements}
The authors thank the anonymous referee for the useful comments, and the IRAM 30m staff for their support in the observations. ACT, PC, and JEP acknowledge the financial support of the European Research Council (ERC; project PALs 320620). The work done by AV and AP was partially supported by the Russian Science Foundation (project 18-12-00351).
\end{acknowledgements}
\bibliography{biblio}
\bibliographystyle{aa}

\begin{appendix}

\clearpage
\onecolumn
\section{Integrated intensity maps}\label{appendix_intensity}
\begin{figure*}[h]
\resizebox{\hsize}{!}{\includegraphics{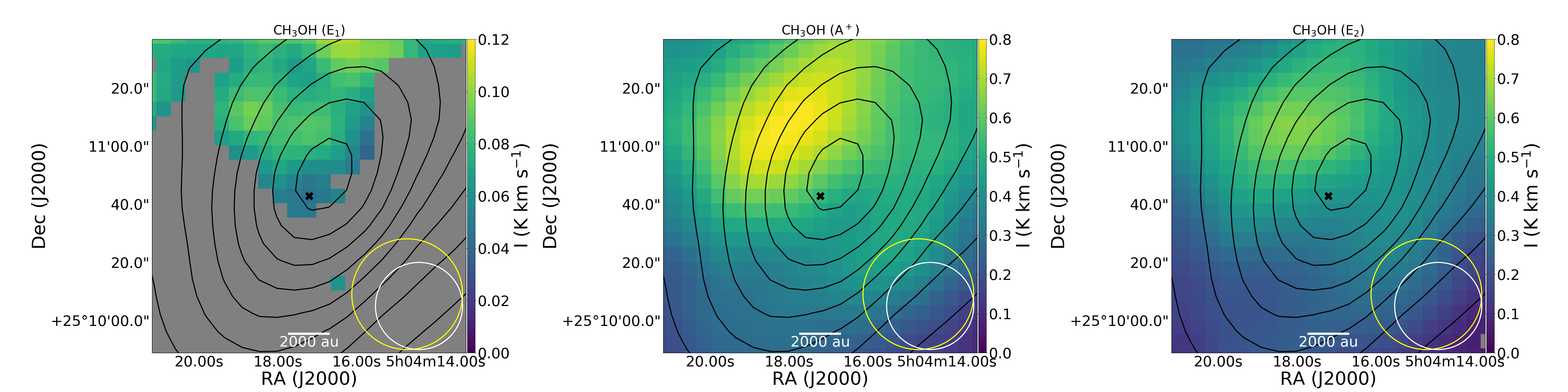}} 
\caption{Integrated intensity maps of the three methanol transitions observed, E$_1$ (\textit{left panel}), A$^+$ (\textit{middle panel}) and E$_2$ (\textit{right panel}). Only pixels with flux values above 3 $\sigma$ detection level are presented. The black contours represent increasing 10\% steps of the N$_{\rm{H}_2}$ column density map, derived by \citetads{2016A&A...592L..11S}. The noise in the integrated intensities is 0.02 K km s$^{-1}$. The HPBWs are shown in the bottom right corner of the figures, in yellow for \textit{Herschel}/SPIRE and in white for the 30m telescope. The black cross marks the dust continuum peak.} 
\label{moment0_ch3oh}
\end{figure*}

\begin{figure}
\centering
{\includegraphics[scale=0.2]{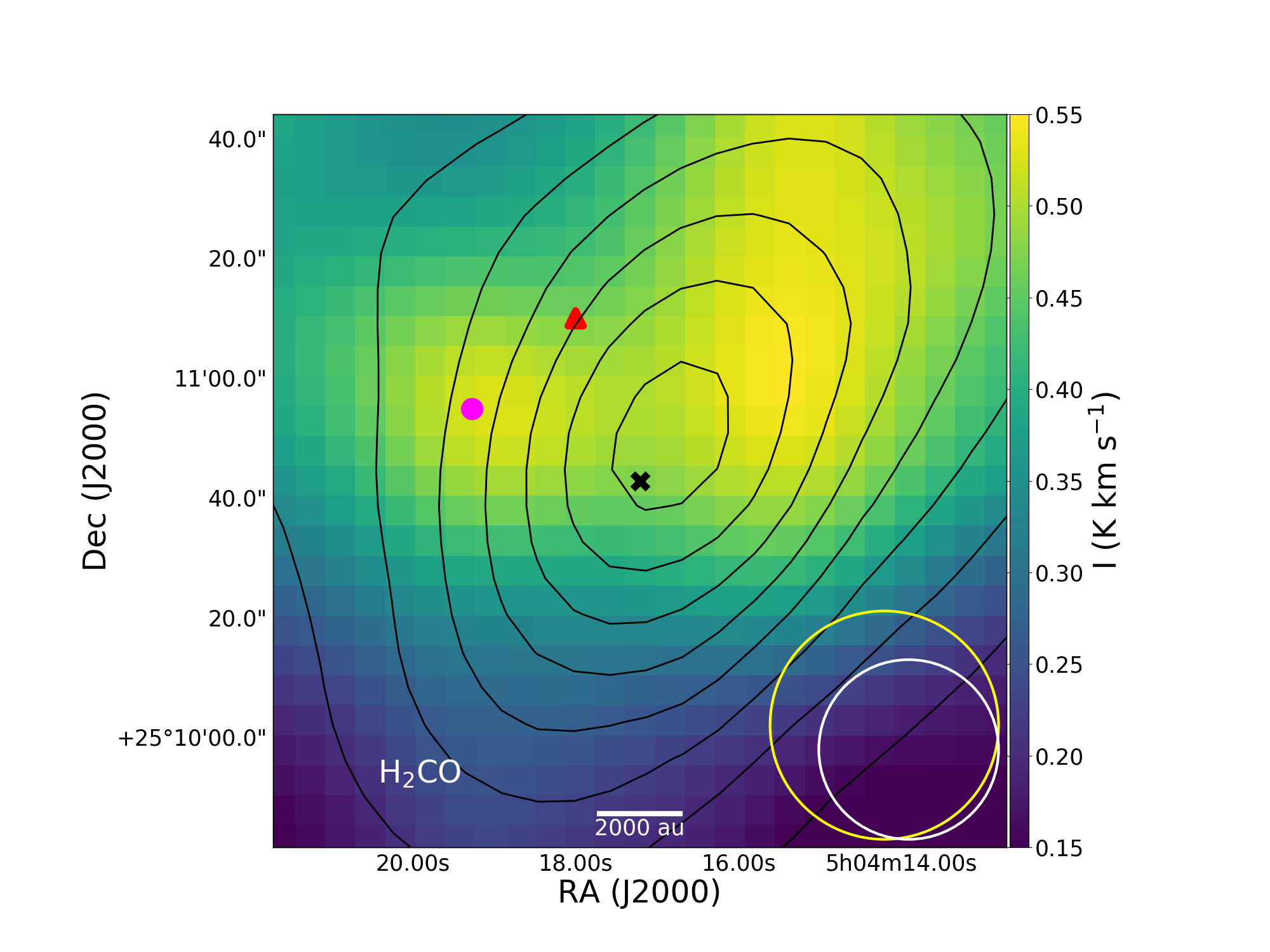}} 
\caption{Integrated intensity map of the H$_{2}$CO (2$_{1,2}$-1$_{1,1}$) line. The black contours represent increasing 10\% steps of the N$_{\rm{H}_2}$ column density map, derived by \citetads{2016A&A...592L..11S}. The error in the integrated intensity is 0.014 K km s$^{-1}$. The HPBWs are shown in the bottom right corner of the figures, in yellow for \textit{Herschel}/SPIRE and in white for the 30m telescope. The black cross marks the dust continuum peak, and the red triangle the methanol peak and the pink circle the C$^{17}$O peak. } 
\label{moment0_h2co}
\end{figure}

%
\clearpage
\twocolumn

\section{Column density maps}\label{appendixb}

\begin{figure}[h]
\resizebox{\hsize}{!}{\includegraphics{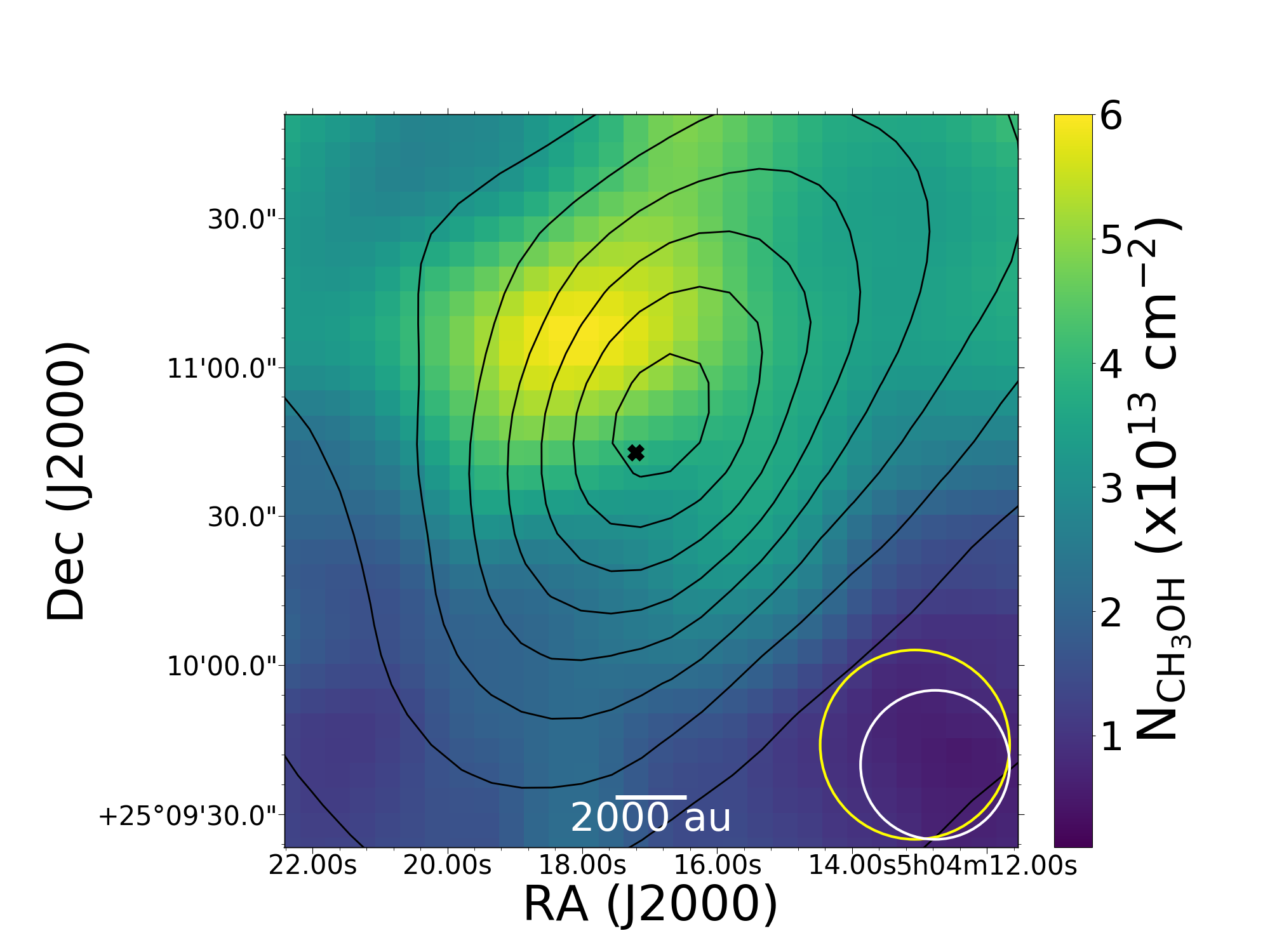}} 
\caption{Column density map of CH$_3$OH derived as explained in Section \ref{cdens}. The black contours represent increasing 10\% steps of the N$_{\rm{H}_2}$ column density map, derived by \citetads{2016A&A...592L..11S}. The HPBWs are shown in the bottom right corner of the figures, in yellow for \textit{Herschel}/SPIRE and in white for the 30m telescope. The black cross marks the dust continuum peak.} 
\label{ch3oh_columndens}
\end{figure}

\begin{figure}[h]
\resizebox{\hsize}{!}{\includegraphics{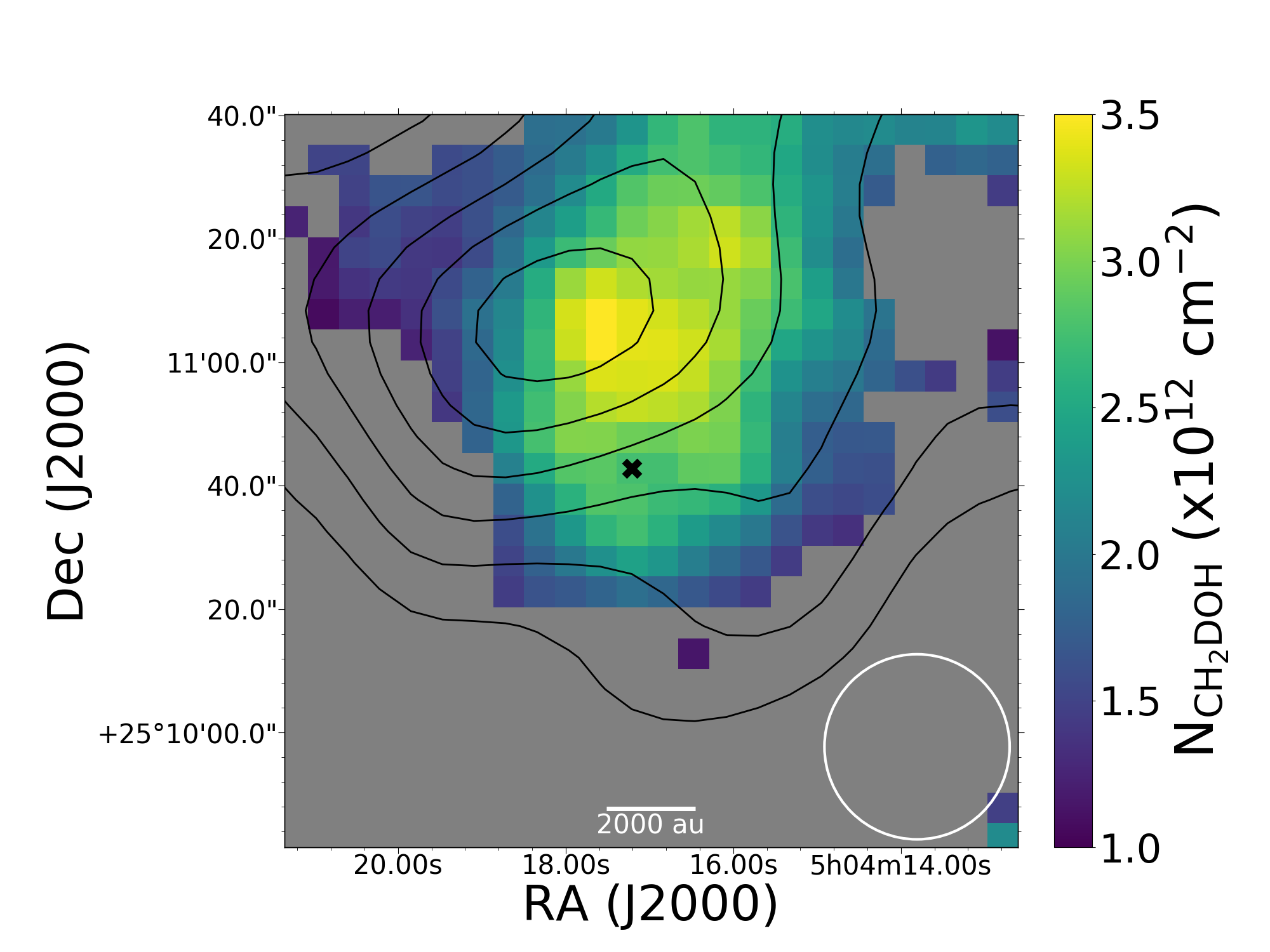}} 
\caption{Column density map of CH$_2$DOH derived as explained in Section \ref{cdens}. The black contours represent 10\% steps in the CH$_3$OH column density. The black cross marks the dust continuum peak. The HPBW is shown in the bottom right corner.} 
\label{ch2doh_columndens}
\end{figure}

\begin{figure}[h]
\resizebox{\hsize}{!}{\includegraphics{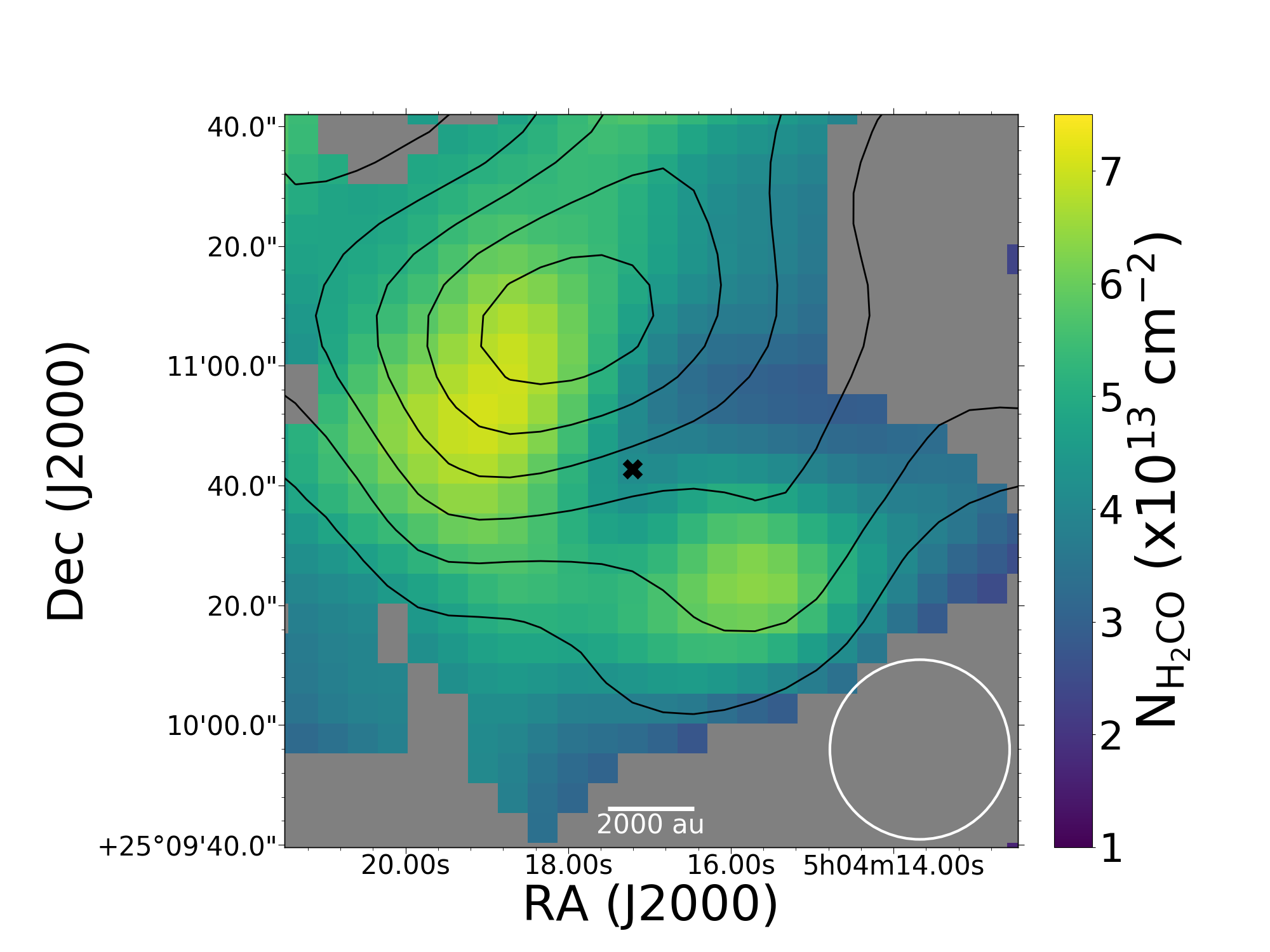}} 
\caption{Column density map of H$_2$CO derived as explained in Section \ref{cdens}. The black contours represent 10\% steps in the CH$_3$OH column density. The black cross marks the dust continuum peak. The HPBW is shown in the bottom right corner.} 
\label{h2co_columndens}
\end{figure}

\begin{figure}[h]
\resizebox{\hsize}{!}{\includegraphics{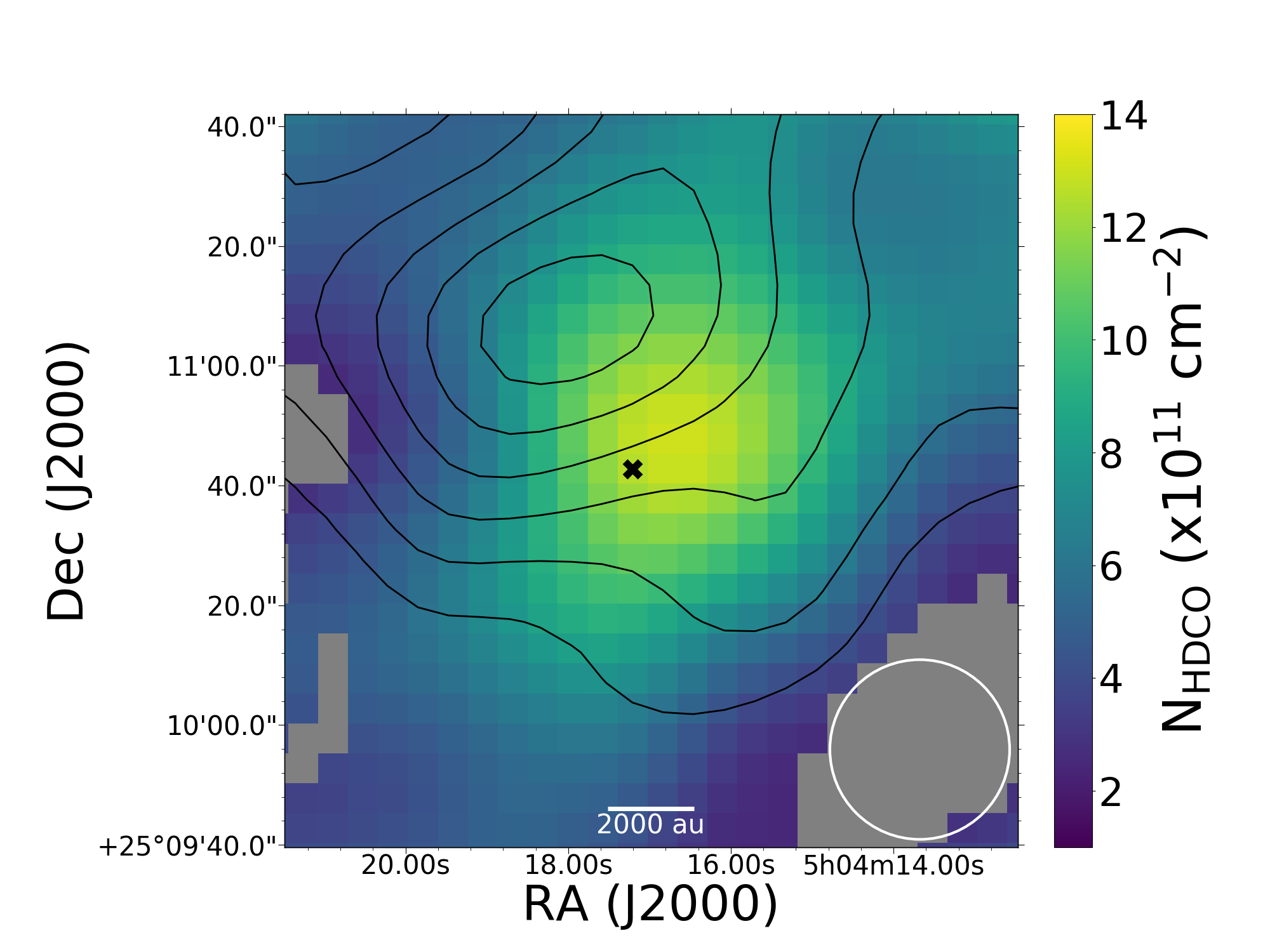}} 
\caption{Column density map of HDCO derived as explained in Section \ref{cdens}. The black contours represent 10\% steps in the CH$_3$OH column density. The black cross marks the dust continuum peak. The HPBW is shown in the bottom right corner.} 
\label{hdco_columndens}
\end{figure}

\begin{figure}[h]
\resizebox{\hsize}{!}{\includegraphics{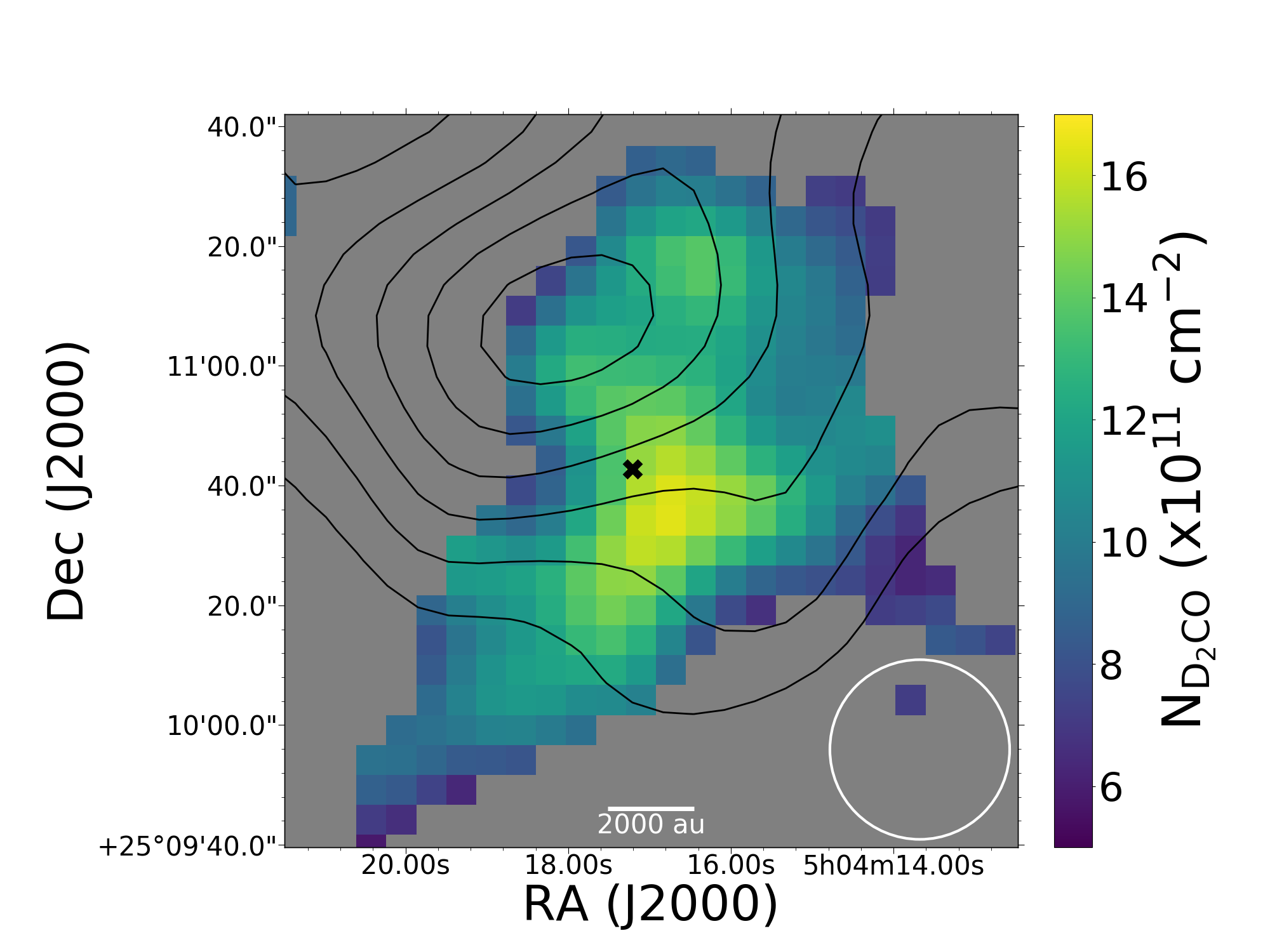}} 
\caption{Column density map of D$_2$CO derived as explained in Section \ref{cdens}. The black contours represent 10\% steps in the CH$_3$OH column density. The black cross marks the dust continuum peak. The HPBW is shown in the bottom right corner.} 
\label{d2co_columndens}
\end{figure}

\begin{figure}[h]
\resizebox{\hsize}{!}{\includegraphics{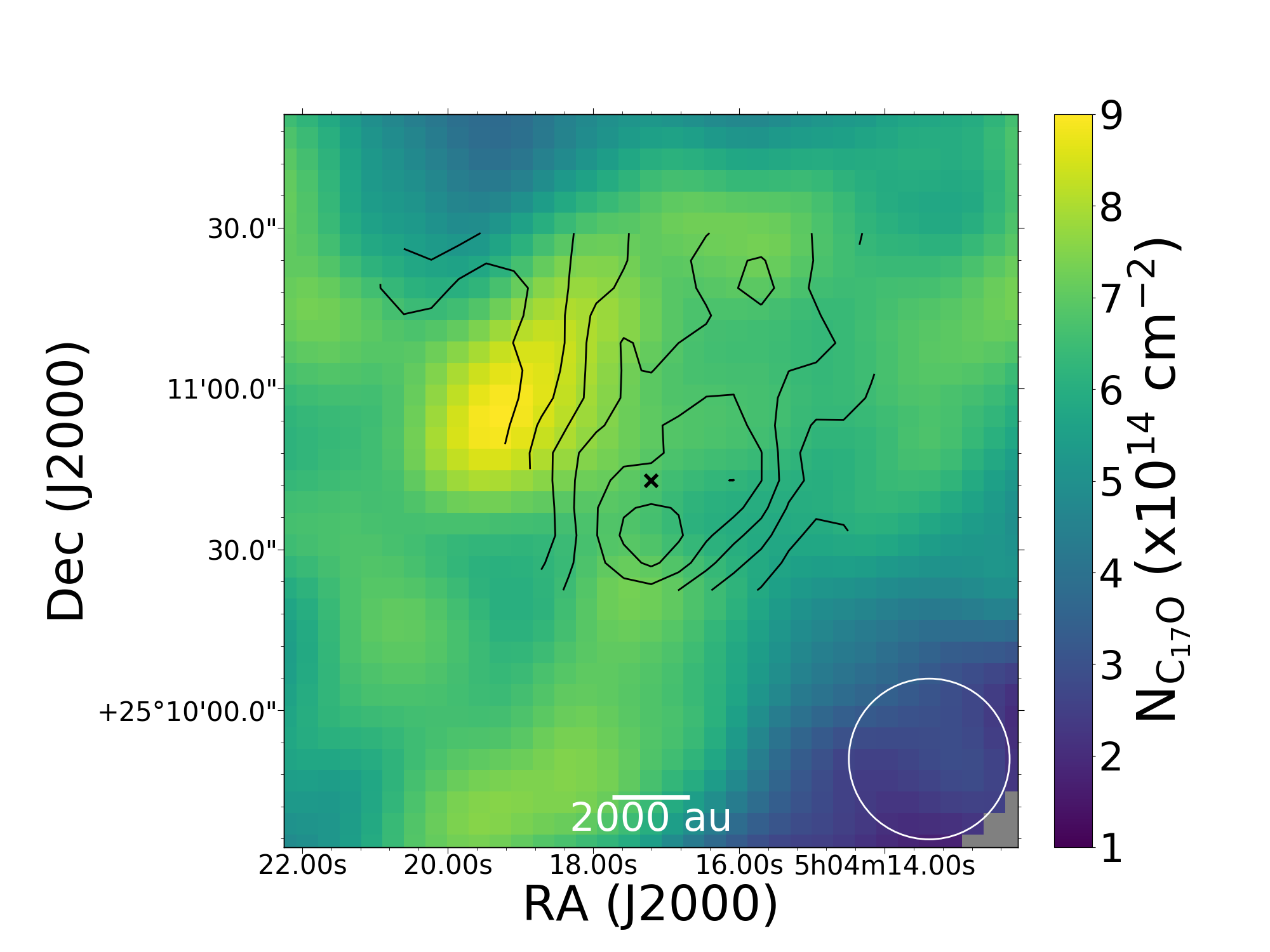}} 
\caption{Column density map of C$^{17}$O derived as explained in Section \ref{cdens}. The black contours represent 10\% steps in the methanol deuterium fraction. The black cross marks the dust continuum peak. The HPBW is shown in the bottom right corner.} 
\label{c17o_columndens}
\end{figure}

\clearpage
\section{Errors on the deuteration maps}
\begin{figure}[h]
\resizebox{\hsize}{!}{\includegraphics{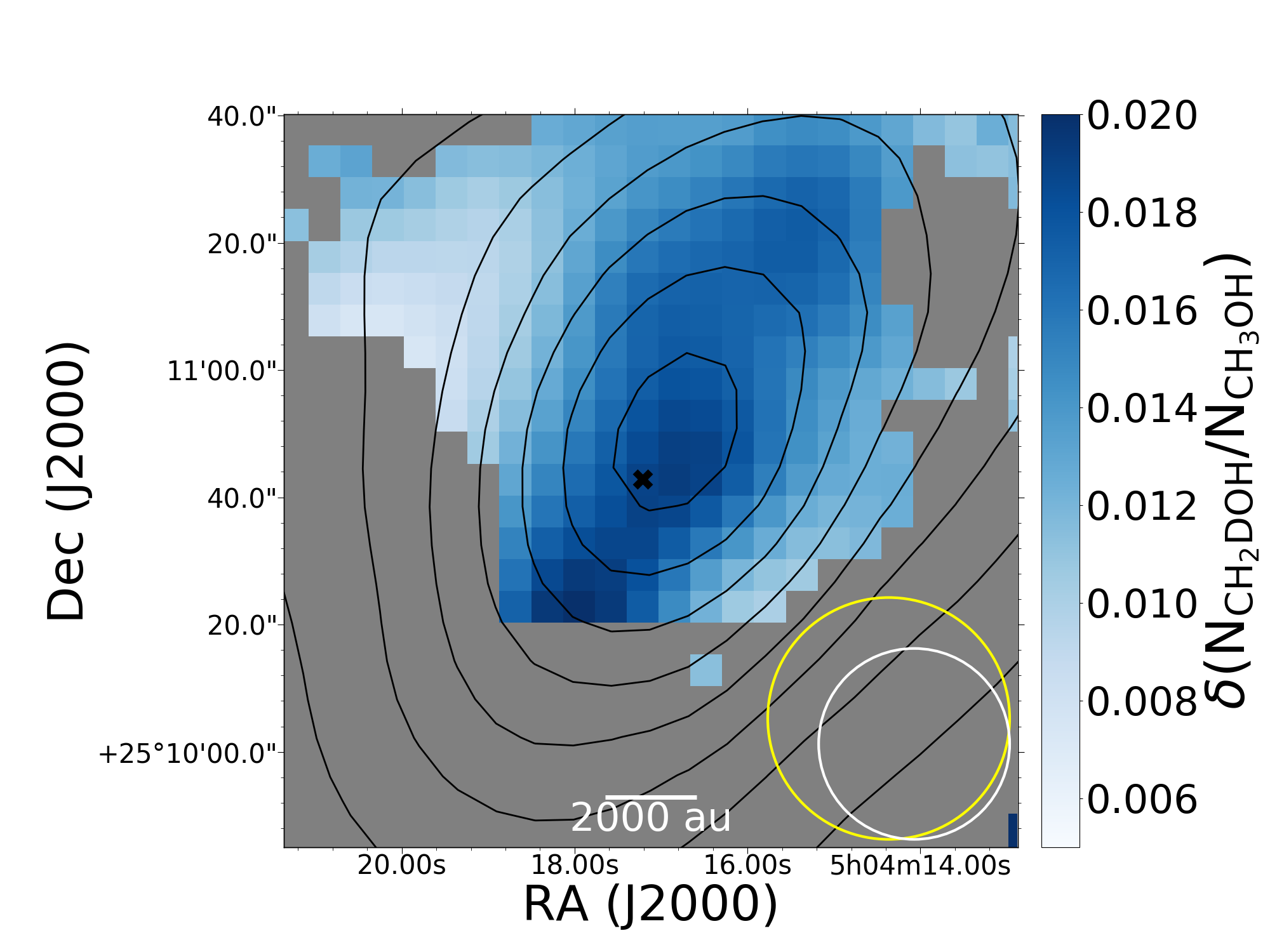}} 
\caption{Error on the ratio between N(CH$_2$DOH) and N(CH$_3$OH). The black contours represent increasing 10\% steps of the peak of the \textit{Herschel} N(H$_2$) map, presented by \SPEt. The HPBWs are shown in the bottom right corner of the figures, in yellow for \textit{Herschel}/SPIRE and in white for the 30m telescope. The black cross marks the dust continuum peak. } 
\label{deut_ch2doh_error}
\end{figure}

\begin{figure*}[h]
\resizebox{\hsize}{!}{\includegraphics{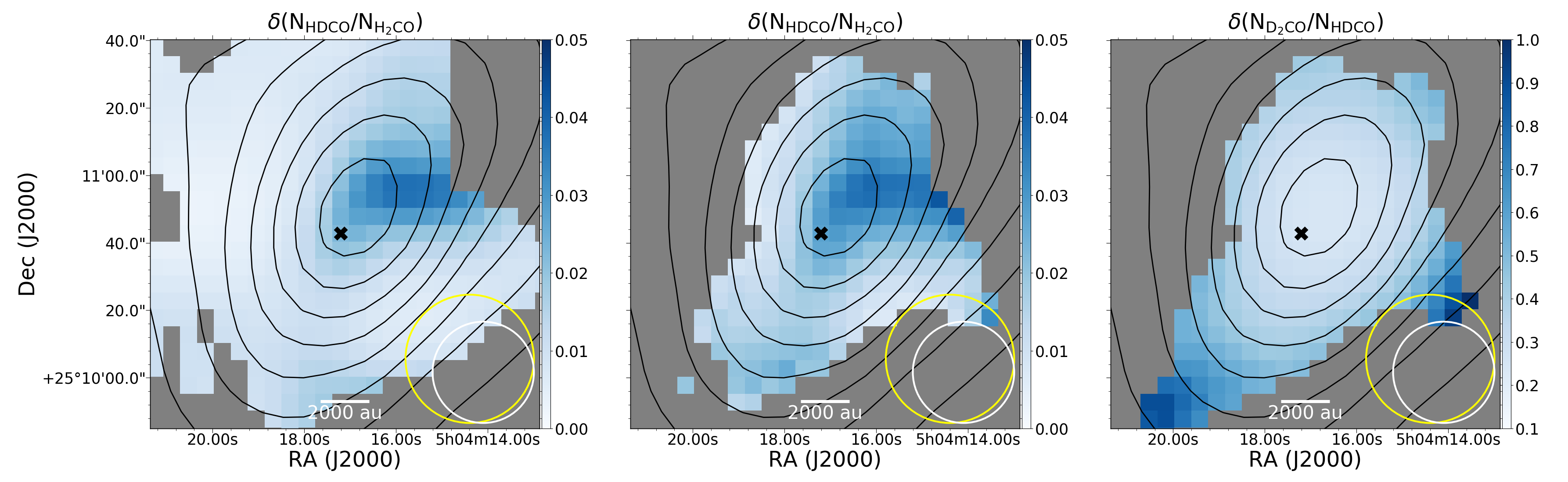}} 
\caption{Error on the ratio between N(HDCO) and N(H$_2$CO) (\textit{left panel}) and N(D$_2$CO) and N(H$_2$CO) (\textit{right panel}). The black contours represent increasing 10\% steps of the peak of the Herschel N(H$_2$) map, presented by \SPEt. The HPBWs are shown in the bottom right corner of the figures, in yellow for \textit{Herschel}/SPIRE and in white for the 30m telescope. The black cross marks the dust continuum peak. } 
\label{deut_h2co_error}
\end{figure*}


\clearpage
\onecolumn

\end{appendix}

\end{document}